\documentclass[%
 aps, rmp,
 letterpaper,floatfix,
 preprint,
 reprint,
 amsmath,amssymb,
 notitlepage,
 longbibliography
]{revtex4-1}

\usepackage{xcolor}
\newcommand{\red}{}
\newcommand{\purple}{}

\newcommand*{\affrugfnd}{Physics of Nanodevices, Zernike Institute for Advanced Materials, University of Groningen, Nijenborgh 4, 9747 AG Groningen, The Netherlands}
\newcommand*{\affnusdept}{Department of Physics, 2 Science Drive 3, National University of Singapore, Singapore 117542, Singapore}
\newcommand*{\affmad}{Imdea Nanociencia, Faraday 9, 28049 Madrid, Spain.}
\newcommand*{\affdon}{Donostia International Physics Center, Paseo Manuel de Lardizabal 4, 20018 San Sebastian, Spain.}

\newcommand*{\affcol}{Department of Physics, Columbia University, New York, New York 10027, USA}
\newcommand*{\affepfl}{Electrical Engineering Institute, Institute of Materials Science and Engineering, \'Ecole Polytechnique F\'ed\'erale de Lausanne (EPFL), Lausanne CH-1015, Switzerland}
\newcommand*{\affman}{Department of Physics and Astronomy, University of Manchester, Manchester, M13 9PL, UK}
\newcommand*{\affngi}{National Graphene Institute, University of Manchester, Manchester M13 9PL, UK}
\usepackage{graphicx}
\usepackage{bm}
\usepackage[breaklinks=true]{hyperref}
\hypersetup{bookmarksnumbered, pdfpagemode=UseOutlines, 
pdfauthor={I.\ J.\ Vera-Marun}, 
pdftitle={Colloquium: Spintronics in graphene and other two-dimensional materials}, pdfdisplaydoctitle, 
colorlinks=true, citecolor=blue, filecolor=blue, linkcolor=blue, urlcolor=blue}

\usepackage{verbatim}
\usepackage{textcomp}

\begin{document}

\title{Colloquium: Spintronics in graphene and other two-dimensional materials}
\author{A.\ Avsar}
\affiliation{\affepfl}
\author{H.\ \surname{Ochoa}}
\affiliation{\affcol}
\author{F.\ \surname{Guinea}}
\affiliation{\affmad}
\affiliation{\affman}
\affiliation{\affdon}
\author{B.\ \surname{\"Ozyilmaz}}
\affiliation{\affnusdept}
\author{B.\ J.\ \surname{van Wees}}
	\email[e-mail: ]{b.j.van.wees@rug.nl} 
	\affiliation{\affrugfnd}
\author{I.\ J.\ \surname{Vera-Marun}}
	\email[e-mail: ]{ivan.veramarun@manchester.ac.uk} 
	\affiliation{\affman}
	\affiliation{\affngi}

\date{\today}

\begin{abstract}
{\purple After the first unequivocal demonstration of spin transport in graphene \cite{tombros_electronic_2007}, surprisingly at room temperature, it was quickly realized that this novel material was relevant for both fundamental spintronics and future applications. Over the decade since, exciting results have made the field of graphene spintronics blossom, and a second generation of studies has extended to new two-dimensional (2D) compounds. This Colloquium reviews recent theoretical and experimental advances on electronic spin transport in graphene and related 2D materials, focusing on emergent phenomena in van der Waals heterostructures and the new perspectives provided by them. These phenomena include proximity-enabled spin-orbit effects, the coupling of electronic spin to light, electrical tunability, and 2D magnetism.}
\end{abstract}

\preprint{Preprint v27a}

\maketitle
\tableofcontents

\section{Introduction}

In 1988, A.\ Fert and P.\ Grunberg independently discovered that the resistance of ferromagnetic/non-magnetic (FM/NM) metallic  multilayers depends on the relative orientation of the FM layers \cite{baibich_giant_1988, binasch_enhanced_1989}. This discovery, termed giant magnetoresistance (GMR), was soon utilized in the magnetic field sensors of hard disk drives \cite{tsang_design_1998}, constituting the first major application of spintronics within modern electronics. The physics behind this achievement has formed the foundation of spin transport experiments and their evolution over the last three decades \cite{fert_nobel_2008}

Spintronics aims to utilize the spin degree of freedom to complement or replace charge as the information carrier for high speed, low-power computing \cite{wolf_spintronics_2006}. The drive for this use of spin-polarized transport started with the proposal of the Datta-Das spin field-effect transistor, which relies on the electrical manipulation of spin information during its propagation in a non-magnetic channel \cite{datta_electronic_1990}. This was followed by other proposals of integrating magnetic semiconductors into spin-based diodes and transistors \cite{flatte_unipolar_2001}, and ultimately all-spin logic circuits with built-in memory \cite{behin-aein_proposal_2010}. These proposals motivated major experimental advances, including the demonstration of spin injection, transport and detection in metals \cite{jedema_electrical_2001} and semiconductors \cite{lou_electrical_2007}. The latter offers the prospect of harnessing the spin degree of freedom for quantum storage and computation \cite{awschalom_semiconductor_2002, flatte_semiconductor_2007}. Despite these initial successes, most of those earlier concepts have not been experimentally realized. The reason for this is that an ideal material platform capable of transporting spin information over long distances at room temperature was still lacking.

Since the discovery of graphene, the field of two-dimensional (2D) materials has become one of the most active areas of research in solid-state physics, mainly because of their prominent mechanical, optical, electrical and magnetic properties \cite{castro_neto_electronic_2009, novoselov_nobel_2011}. High electronic charge mobility, low spin orbit coupling strength, negligible hyperfine interaction and gate tunability, are part of the properties which have established graphene also as an emerging material for spintronics \cite{tombros_electronic_2007}. Graphene exhibits the longest spin relaxation length ever measured at room temperature \cite{ingla-aynes_24_2015, drogeler_spin_2016}. 
{\red This property enables the transmission and manipulation of spin signals within complex multi-terminal device architectures \cite{kamalakar_long_2015, gebeyehu_spin_2019}.}
Recently, the field has moved beyond graphene towards exploring the properties of other 2D crystals and their heterostructures, which serve as testbeds for inducing new emergent functionalities. 
{\red The observation of long spin relaxation lengths \cite{drogeler_spin_2016} together with spin-charge conversion \cite{safeer_room-temperature_2019} and spin manipulation \cite{avsar_gate-tunable_2017} capabilities in van der Waals heterostructures make 2D crystals appealing material systems for the development of low-power spintronics devices. The latter include, among others, devices based on tunnel magnetoresistance \cite{kim_one_2018}, spin-transfer torque \cite{lin_spin_2013}, and spin logic gates \cite{wen_experimental_2016}}.

Given the increasing effort on the search for new spintronics phenomena in 2D materials, there is a present need for a review and critical discussion of recent experimental and theoretical progress within the field of 2D spintronics. We note there are relevant reviews on spin transport \cite{meservey_spin-polarized_1994, zutic_spintronics:_2004, chappert_emergence_2007, wolf_spintronics_2006}, graphene \cite{castro_neto_electronic_2009, peres_colloquium:_2010, das_sarma_electronic_2011}, and graphene spintronics \cite{pesin_spintronics_2012, shiraishi_electrically-generated_2012, han_spin_2012, seneor_spintronics_2012, roche_graphene_2014, gurram_electrical_2018, garcia_spin_2018, roche_graphene_2015, zutic_proximitized_2019}. 
Our aim in this review is to offer a broad and balanced coverage, starting with an introduction to fundamentals for those new to the fields of spintronics and graphene, while encompassing critical discussions that address recent advances and point to challenges and future directions in this vibrating field. 
{\red While brief discussions on chemically modified graphene and alternative spin injection schemes will be also provided, this colloquium primarily focuses on electronic spin transport in pristine 2D materials and novel devices based on their heterostructures}.

\subsection{Initial spin transport experiments}

At the heart of most spin transport phenomena is the creation of spin accumulation in a NM material. The latter can be electrically induced by applying a charge current to the NM via a FM contact. The injected spin information propagates throughout the NM as a spin current, which can be detected using a second FM contact. Such detection is possible as long as the separation between the injector and detector contacts is less than the characteristic length on which the non-equilibrium spin information relaxes. 
Within these conditions, the resistance of the device depends on the relative magnetization direction of the injector and detector contacts, which is detected as a resistance switch when their relative orientation changes from parallel to the antiparallel \cite{johnson_interfacial_1985, jedema_electrical_2001}. Such a spin valve signal in FM/NM/FM structures is therefore utilized for the realization of spin injection, spin transport and spin detection. In this local architecture, the charge and spin currents are co-located within the same NM channel. The measured signal is therefore a combination of a spin-dependent resistance, representative of the spin transport, and a spin-independent resistance, purely associated to the charge transport. Since the ratio of the spin-dependent resistance to the spin-independent resistance is generally small, and 
charge-based phenomena can mimic the spin signal under study, this two-terminal measurement geometry is not ideal for probing spin-dependent transport.

\begin{figure}
\includegraphics[width=1.0\columnwidth]{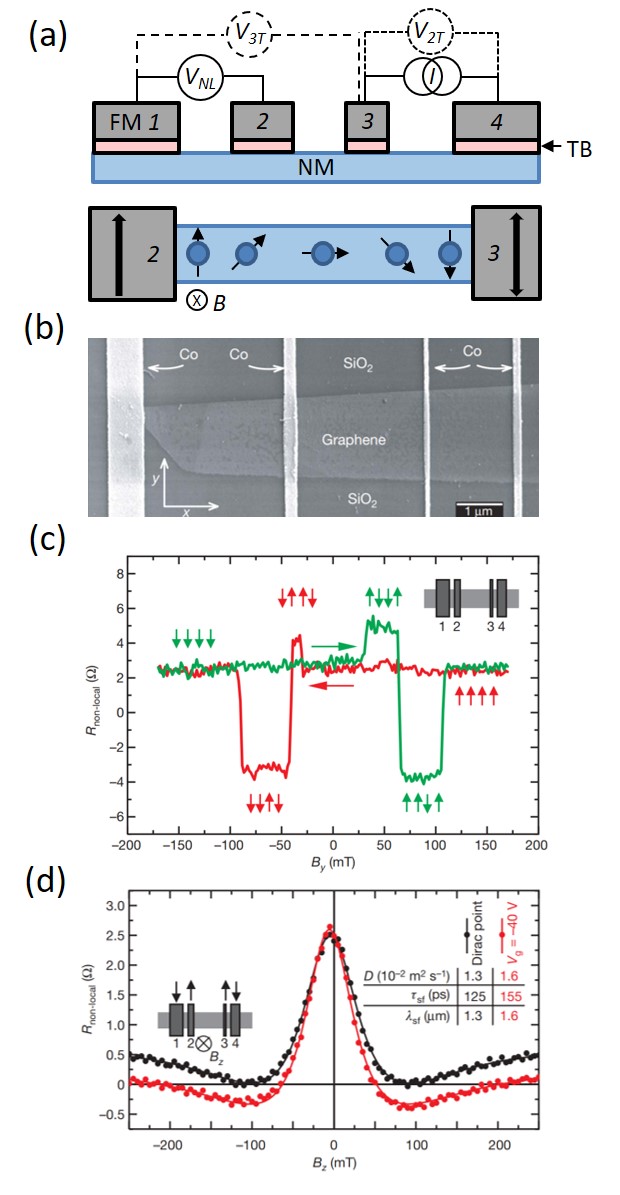}
\caption{
\emph{Measurement configurations and spin transport in graphene.}
(a) Four-terminal nonlocal ($V_\text{NL}$), three-terminal Hanle ($V_\text{3T}$) and two-terminal local ($V_\text{2T}$) measurement geometries. Ferromagnetic (FM) contacts are used to inject and detect spins in a non-magnetic (NM) channel via a tunnel barrier (TB). The bottom cartoon illustrates a top-view of spin precession under an out-of-plane applied magnetic field ($B$). (b) Scanning electron micrograph of the first non-local graphene spin valve device. 
(c) Non-local spin valve signal in monolayer graphene. Vertical arrows represent the polarization direction of the contacts, lateral arrows represent the magnetic field sweep direction. (d) Corresponding Hanle spin precession signal, for parallel orientation of the contacts. Adapted from \citet{tombros_electronic_2007}.
}
\label{fig:intro}
\end{figure}

Multi-terminal measurement geometries are thus favoured for the unambiguous study of spin-dependent transport (Fig.~\ref{fig:intro}). 
One alternative is the three-terminal Hanle geometry, where a single magnetic contact is used to create and detect spin accumulation. The latter is evidenced by the modulation of the contact resistance due to the spin precession effect of an applied magnetic field. 
This measurement configuration has advantages for probing spin lifetimes in highly-resistive channel materials or in materials with a short spin relaxation length, and has been utilized to detect spin injection in GaAs \cite{lou_electrical_2006, lou_electrical_2007}, Si \cite{dash_electrical_2009} and Ge \cite{jeon_electrical_2011}. However, a disadvantage of this configuration is that the same contact is used to electrically inject and detect the spin information, which does not fully solve the issue of mixing charge and spin phenomena, as evidenced in measurements showing spin accumulations that are inconsistent with the theoretical expectations \cite{tran_enhancement_2009}. Therefore, the ideal method is that of a four-terminal non-local geometry, see Fig.~\ref{fig:intro}. 
This non-local geometry has been widely utilized to spatially separate the paths of the charge current and the detected spin current, providing the reliable measurement of spin-dependent transport. This method relies on the detection of pure spin current and has been first established for metallic spin valves \cite{johnson_interfacial_1985, jedema_electrical_2001} and later also applied to semiconductor-based devices \cite{lou_electrical_2007}.

{\red Quantitative extraction of the parameters that control spin transport in a diffusive channel, namely the spin relaxation time $\tau_s$ and the spin diffusion coefficient $D_s$, can be achieved by measuring the dependence of the spin signal on the separation between injector and detector electrodes. The observed decay with increasing separation yields the spin relaxation length, $\lambda_s = \sqrt{D_S \tau_s}$, while $D_s$ can be calculated from the resistivity via the Einstein relation \cite{jedema_electrical_2001}. Nevertheless, this requires the study of multiple devices with different separation and is sensitive to the reproducibility of other parameters, like the spin injection efficiency of the contacts, which also control the magnitude of the spin signal. Hanle spin precession is an alternative measurement based on studying the dependence of the spin signal on an applied magnetic field $B$ perpendicular to the orientation of the injected spin \cite{jedema_electrical_2002}. Here, a torque is exerted on the electron spin, which is forced to precess at the Larmor frequency, as sketched in Fig.~\ref{fig:intro}(a). In a fully coherent 1D channel the spin precession would lead to an oscillatory response as a function of $B$. Nevertheless, in a diffusive 2D or 3D channel each electron spin follows a different path towards the detector, due to momentum scattering, with each one having a different transit time and therefore a different precession angle, causing dephasing in the net spin accumulation. Furthermore the spin accumulation relaxes back to its equilibrium value. The combined effect of spin dephasing and relaxation, controlled by $D_s$ and $\tau_s$, determine the response of the spin signal as a function of $B$, as shown in Fig.~\ref{fig:intro}(d). This powerful technique removes the requirement to study multiple devices to extract spin parameters, as it relies on fitting the lineshape of the response only based on the two fundamental parameters controlling spin transport \cite{jedema_apl}. Hanle spin precession is the golden standard for the unambiguous demonstration of spin transport using a four-terminal non-local geometry and it has been extensively applied to metallic \cite{jedema_electrical_2002}, semiconducting \cite{lou_electrical_2007} and graphene \cite{popinciuc_electronic_2009, maassen_contact-induced_2012} spin channels.}

Spin transport experiments in graphene have utilized the measurement techniques described above. The first report on spin transport demonstrated a hysteretic magneto-transport response with a 10$\%$ change in resistance by utilizing a two-terminal local geometry \cite{hill_graphene_2006}.  Shortly after, \citet{tombros_electronic_2007} unequivocally demonstrated room-temperature spin transport with micrometer-long spin relaxation lengths in monolayer graphene, by employing a tunnel barrier for efficient spin injection and measuring in a four-terminal non-local geometry. This measurement configuration then became the standard method to characterize the distinct spin transport properties of graphene, as demonstrated by \citet{cho_gate-tunable_2007} and \citet{popinciuc_electronic_2009} who modulated spin transport in graphene via the application of \emph{vertical} electric fields. On the other hand, application of \emph{lateral} electric fields also demonstrated to produce spin drift and therefore modulate spin injection efficiencies \cite{jozsa_electronic_2008, jozsa_controlling_2009}. 

After the characterization of basic spin transport properties of graphene fabricated on standard Si/SiO$_2$ substrates, \citet{jozsa_linear_2009} examined the dominant spin relaxation mechanism for the first time by studying the relation between momentum and spin relaxation times in monolayer graphene. A further discussion on spin relaxation is presented in Section \ref{sec:relaxation}. Initial devices also demonstrated the capability to achieve robust spin polarization in mono and multi-layer graphene, by demonstrating a linear dependence of non-local voltage on injected current up to 10 mA \cite{muramoto_analysis_2009, shiraishi_robustness_2009}. Such superior spin injection properties, in comparison with previous semiconductor-based devices, is a result of the transport properties of graphene and the suppression of any significant interface spin scattering, as later confirmed by the observation of identical spin transport parameters in graphene spin valve devices measured with either three and four-terminal techniques over a wide range of temperatures \cite{dankert_spin_2014}.

\subsection{Fundamentals of spin-orbit coupling in 2D crystals}

In the solid state electrons usually move much slower than light, with a speed $v\ll c$. However, relativistic corrections are not completely negligible, representing both a major limitation to spin transport and a source of opportunities for spin injection, manipulation and detection. The leading correction in $v/c$ is provided by the Hamiltonian 
\begin{align}
\mathcal{H}_{SO}=\frac{1}{2m^2c^2}\left(\bm{\nabla} V\times\mathbf{p}\right)\cdot\mathbf{s},
\label{eq:so_general}
\end{align}
where $m$ is the electron mass and $\mathbf{p}$ and $\mathbf{s}=\left(s_x,s_y,s_z\right)$ represent the linear momentum and spin operators, respectively. In a solid, $V$ is just the crystalline potential. 
{\red As a first approximation, we can consider an atomic insulator, where electrons are bound to the nuclei of the constituents of the solid by a hydrogen-like potential of the form  $V(r)\propto Z/r$, $Z$ being the atomic number. Eq. ~\eqref{eq:so_general} can be written as $\mathcal{H}_{SO}=\Delta_{SO}\mbox{ }\mathbf{L}\cdot\mathbf{s}$, where $\mathbf{L}=\mathbf{r}\times\mathbf{p}$ is the angular momentum operator and $\Delta_{SO}\propto Z/r^3$. The averaged \textit{intra-atomic} spin-orbit coupling (SOC) is dominated by electrons close to the nucleus, at distances of the order of the Bohr radius, $r\sim a_B\propto Z^{-1}$, for which the nuclear field remains almost unscreened. Since the probability of finding an electron near the nucleus scales as $\sim Z^{-2}$ \cite{Landau_book}, the intra-atomic SOC constant goes like $\Delta_{SO}\propto Z^2$: the heavier the atom, the larger the relativistic effect. In fact, the fast decay of the spin-orbit interaction with the distance to the nucleus justifies a tight-binding description. For the same reasons, relativistic corrections are smaller for bands built up from higher atomic orbitals.} 
{\purple Such $Z^2$ dependence of SOC in elemental 2D materials has recently been demonstrated by \citet{kurpas_spin-orbit_2019}.}

The spin quantum number reflects the way that the electronic wave function transforms under rotations. Hence, the point group of the lattice (the set of proper and improper rotations that leave invariant the crystal structure) has a strong influence on the relativistic corrections to the electronic bands. The basis of Bloch states around high-symmetry points of the Brillouin zone must be adapted to the irreducible representations of the so-called \textit{double group} \cite{Dresselhaus}, consisting of the original symmetry operations plus the rotation by $2\pi$ along one of the principal axis of the crystal that defines the natural spin quantization. For 2D materials, this is usually the axis perpendicular to the plane of the crystal. The bands are described in terms of the direct product $\Gamma\times D_{1/2}$, where $\Gamma$ labels the irreducible representation associated with the orbital part of the wave function, and $D_{1/2}$ is the spinor representation generated by Pauli matrices $s_i$. This group-theory approach can be simplified in practice. Typically, the strength of the spin-orbit interaction is much weaker than the energy separation between states labeled by different angular-momentum numbers. Therefore, the main orbital character of the bands does not change much, and the indices associated with the irreducible representations of the original point group are still good quantum numbers.

Improper rotations play a crucial role given the pseudo-vectorial nature of the spin operator. If the crystal structure possesses a complete center of inversion, then the original two-fold degeneracy of the bands is preserved. Otherwise, the bands are spin-split except at time-reversal symmetric points. Nevertheless, a plane of inversion always protects the perpendicular projection of the spin close to high-symmetry points. Graphene is mirror symmetric, meaning that $s_z$ is still a good quantum number in the low-energy limit even in the presence of relativistic interactions. Moreover, in non-centrosymmetric materials with mirror symmetry, like transition-metal dichalcogenides (TMDC), the large spin-splittings of the bands provide a strong protection of $s_z$ against external perturbations. 

Despite the small atomic number of carbon, the effect of relativistic interactions in the spectrum of its different allotropes is a subject of intensive research. The spin-orbit coupling in graphite was first analyzed in the pioneering works by Slonczewski \cite{Slonczewski_Weiss} and Dresselhaus \cite{Dresselhaus_bis}. 
In the next section we focus on single and few layer graphene allotropes and TMDCs, describing how SOC can be exploited in spintronic devices.

\section{State of the art in graphene spintronics}

\subsection{Spin-orbit coupling in graphene}
 

Graphene consists of a single layer of $sp^2$ hybridized carbon atoms \cite{castro_neto_electronic_2009,Katsnelson_rev}. In-plane $p_{x,y}$ and $s$ orbitals participate in the strong $\sigma$-bonds that keep carbon atoms covalently attached, forming a trigonal planar structure with a distance between atoms of $a=1.42$ \AA. The remaining electron occupying the $p_z$ orbital is free to hop between neighboring sites, leading to the $\pi$ bands responsible for the conductive properties. The low-energy $\pi$ bands form Dirac cones at the two inequivalent corners of the hexagonal Brillouin zone. The effective Hamiltonian around these points, including relativistic corrections, reads 
\begin{align}
\mathcal{H}= \,& \hbar v_F\left(\tau_z\sigma_x k_x+\sigma_y k_y\right)+\Delta_{KM}\tau_z\sigma_zs_z\nonumber\\
& +\Delta_{BR}\left(\tau_z\sigma_x s_y-\sigma_y s_x\right),
\label{eq:Dirac}
\end{align}
where $v_F\approx c/100$ is the Fermi velocity Dirac electrons and the operators $\sigma_{x,y,z}$ and $\tau_{x,y,z}$ are Pauli matrices acting on the sublattice and valley degrees of freedom of the wave function, respectively. The second term in Eq.~\eqref{eq:Dirac} corresponds to the intrinsic spin-orbit coupling \cite{Kane_Mele1}. It is fully compatible with both hexagonal ($C_{6v}$ point group) and mirror symmetries of free-standing graphene. If this latter symmetry is broken, e.g., by the presence of a substrate, then a Bychkov-Rashba coupling is also generated. The spin-orbit couplings in confined geometries have been analyzed in~\citet{Zarea}, \citet{Lopez-Sancho_etal}, \citet{Lenz2013}, and \citet{Santos_etal}.

\begin{figure}
\includegraphics[width=1.0\columnwidth]{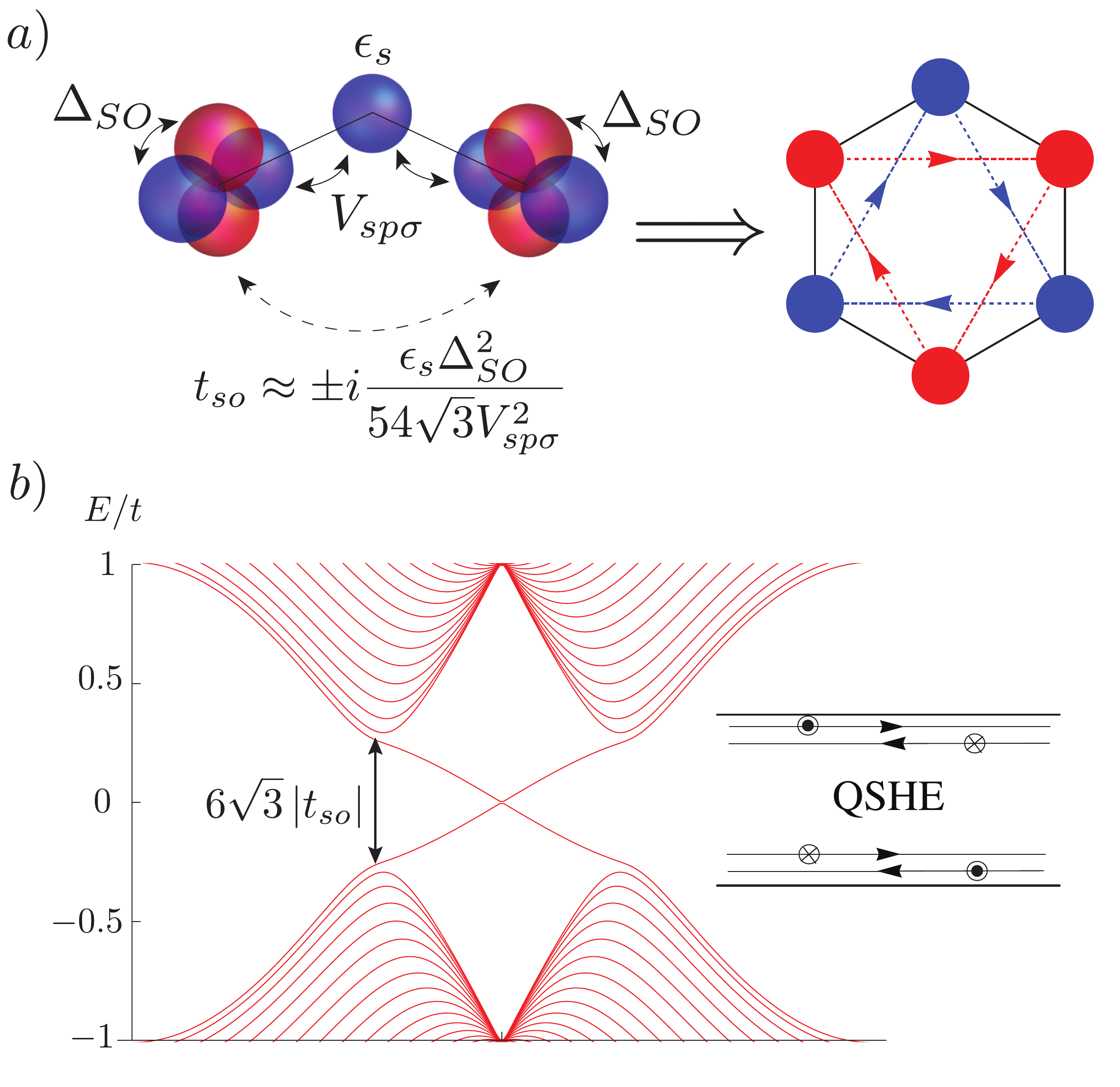}
\caption{a) Spin-dependent effective hopping $t_{so}=\Delta_{KM}/3\sqrt{3}$ mediated by $\sigma$ orbitals (in blue). The upper/lower sign applies to spin up/down electrons, with the hooping defined along the arrows in the right panel. b) Electronic spectrum of a zig-zag ribbon of 30 unit-cells width ($t_{so}=0.1t$, where $t$ represents the nearest-neighbor hopping). The inset shows a sketch of the sub-gap counter-propagating modes localized at the edges.}
\label{fig:qshe}
\end{figure}

The minimal microscopic model that captures spin-orbit effects should include at least both $\pi$ and $\sigma$ orbitals. 
{\red In this model, the intrinsic SOC arises from virtual transitions into $\sigma$ states mediated by the intra-atomic spin-orbit interaction, as represented in Fig.~\ref{fig:qshe}. Second-order perturbation theory gives} \citep{Huertas_etal,Min_etal,Konschuh_etal} 
{\red 
\begin{align}
\label{eq:estimate}
\Delta_{KM}=\frac{\epsilon_s \Delta_{SO}^2}{18 V_{sp\sigma}^2}\approx1\mbox{ \textmu eV},
\end{align}
where $V_{sp\sigma}\approx 4.2$ eV represents the hopping between $p_{x,y}$ and $s$ orbitals, $\epsilon_s\approx-7.3$ eV is the energy of the latter measured with respect to the intrinsic Fermi level \cite{Tomanek_Louie} and the intra-atomic spin-orbit coupling for carbon is $\Delta_{SO}\approx7.86$ meV. This second-order effect is exceeded by the contribution from the hybridization with higher-energy $d$ orbitals \citep{Slonczewski_Weiss}, leading to $\Delta_{KM}\approx 12$  \textmu eV \cite{Yao_etal,Gmitra_etal,Abdelouahed_etal,Konschuh_etal}. Electron-electron interactions \cite{Kane_Mele1} and the coupling with flexural optical phonons can also contribute to this coupling \cite{Ochoa_etal1}. Recent electron-spin resonance measurements give $\Delta_{KM}\approx 21$ \textmu eV \cite{Sichau_etal}.}

\begin{figure}
\includegraphics[width=1.0\columnwidth]{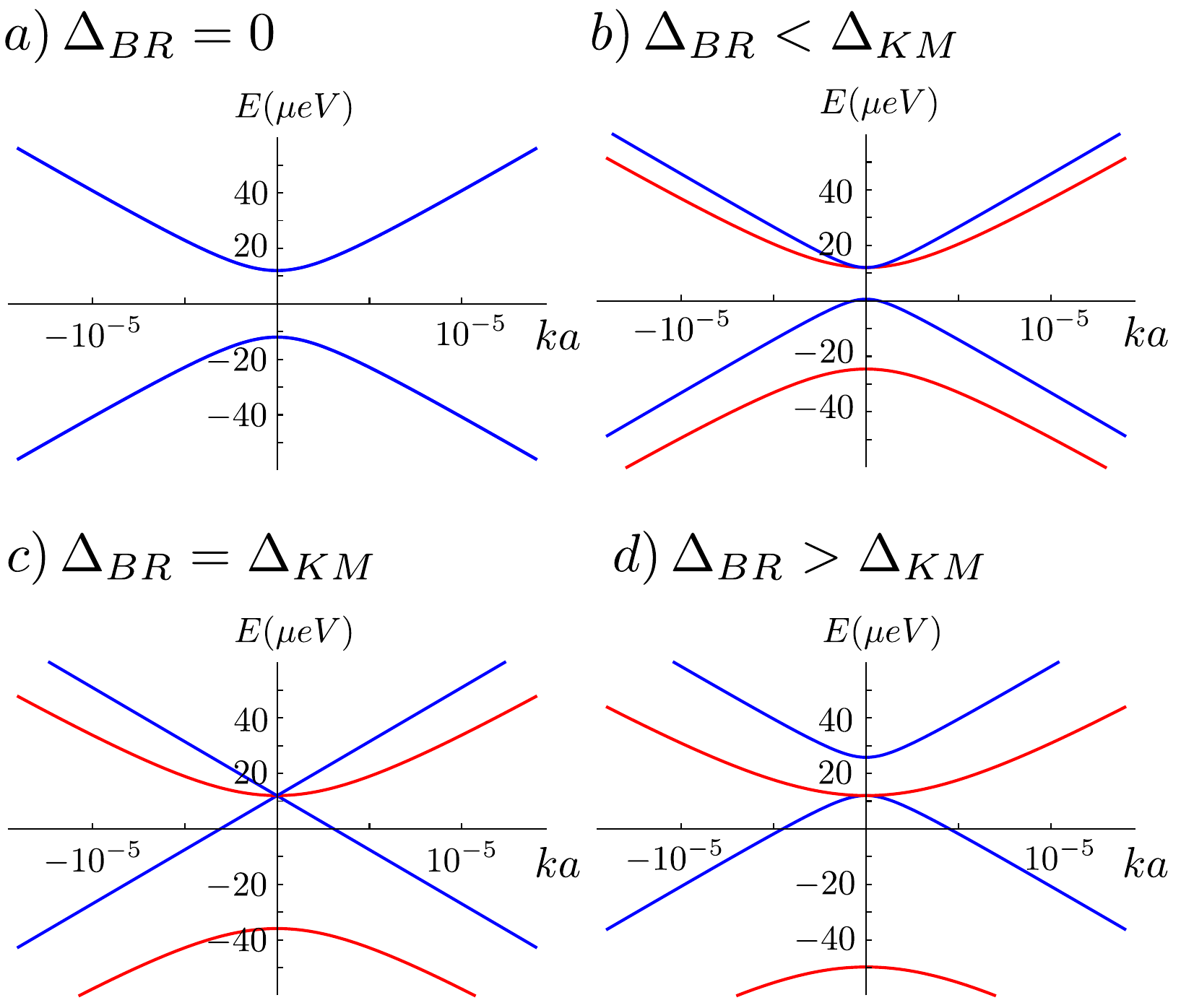}
\caption{
{\red Low-energy bands of graphene around the corners of the Brillouin zone (all values in \textmu eV): a) $\Delta_{BR}=0$, b) $\Delta_{BR}=6$, c) $\Delta_{BR}=12$, and d) $\Delta_{BR}=18$ ($\Delta_{KM}=12$ in all cases). Note that the Bychkov-Rashba coupling lifts the spin degeneracy of the bands. Blue and red colors represent opposite helicities (the approximated spin polarization of Bloch electrons lie within the graphene plane along an axis orthogonal to their crystal momenta).}}
\label{fig:bands_graphene}
\end{figure}

The Bychkov-Rashba term in the second line of Eq.~\eqref{eq:Dirac} removes the spin degeneracy of the bands and tends to close the intrinsic spin-orbit gap. As we mentioned before, it is naturally present when the mirror symmetry is broken while the planar $C_{6v}$ symmetry is preserved. A handy example is the case of an electric field perpendicular to the graphene sample. 
{\red A dipolar coupling of the form $\mathcal{H}_{d}=e\,\mathcal{E}_z\cdot z$ induces transitions between $p_z$ and $s$ orbitals parametrized by $\zeta\equiv\left\langle s|z|p_z\right\rangle$. The electron goes back to the $\pi$ band through the intra-atomic spin-orbit interaction, flipping the spin. Perturbation theory gives \citep{Huertas_etal,Min_etal,Konschuh_etal}
\begin{equation}
\Delta_{BR}=\frac{e\mathcal{E}_z\zeta\Delta_{SO}}{3V_{sp\sigma}}\approx 0.1 \, \text{meV} \times \mathcal{E}_z\left[\text{V}/\text{nm}\right],
\end{equation}
where we have taken $\zeta=3a_B$ for the numerical estimation. \textit{Ab initio} calculations lower this estimate by an order of magnitude \citep{Gmitra_etal,Konschuh_etal}. Fig.~\ref{fig:bands_graphene} shows the different band topologies resulting from the competition between the intrinsic and Bychkov-Rashba SOC terms.}

\subsubsection{hBN-Gr: Long-distance spin transport}

\begin{figure*}
\includegraphics[width=1.0\textwidth]{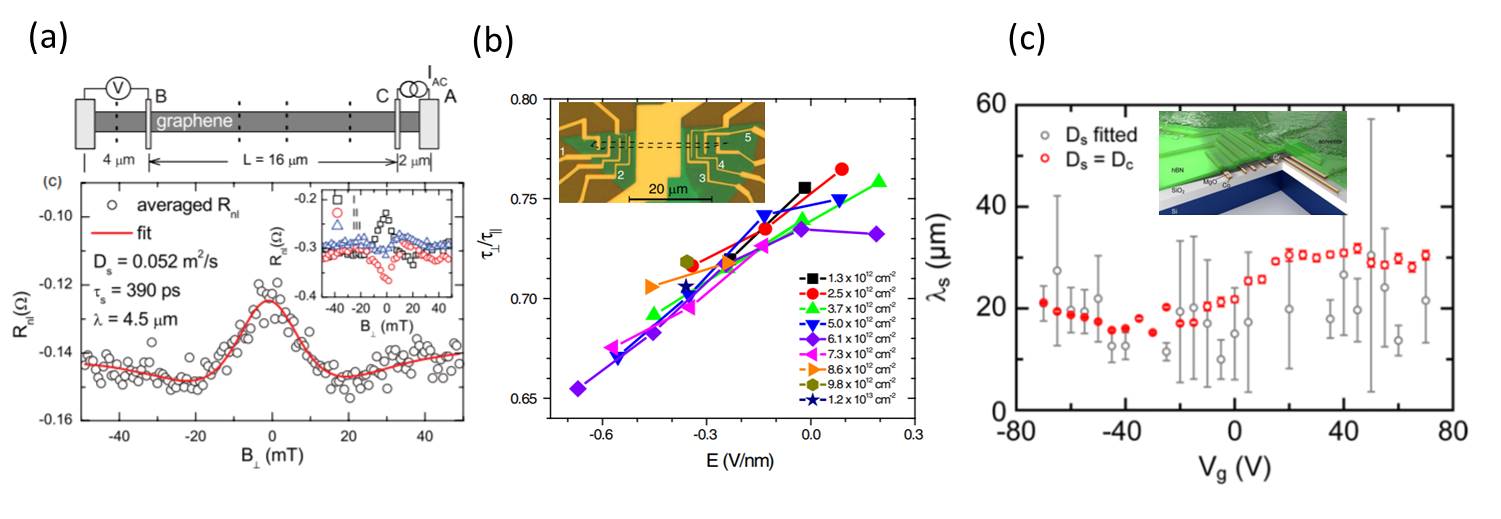}
\caption{
\emph{High Quality Heterostructures for Spin Transport.}
(a) Hanle spin precession measurement for a distance between injector and detector of 18~\textmu m. 
From \citet{zomer_long-distance_2012}. 
(b) Electric field dependence of the ratio in- and out-of-plane injected spins at fixed carrier concentrations. Inset shows the optical picture of a completed device. From \citet{guimaraes_controlling_2014}. 
(c) Back gate voltage dependence of spin relaxation length in an inverted graphene spin valve. Inset shows the device schematics. From \citet{drogeler_spin_2016}.
}
\label{fig:highQ}
\end{figure*}

As a result of the small intrinsic spin-orbit coupling it was expected that graphene would exhibit a long-distance spin transport. This is evident when combining graphene with hexagonal boron nitride (hBN). Due to hBN's reduced trapped charge concentration and atomically flat surface compared to conventional SiO$_2$, hBN has been employed as an ideal substrate for boosting the electronic charge quality of graphene
\cite{dean_boron_2010}. Recently, hBN has been adapted in graphene spin valves as substrate 
\cite{zomer_long-distance_2012}, encapsulating layer \cite{guimaraes_controlling_2014, ingla-aynes_24_2015, avsar_electronic_2016, gurram_spin_2016} and tunnel barrier \cite{yamaguchi_electrical_2013, kamalakar_enhanced_2014, singh_nanosecond_2016} to improve its spin transport properties and realize new device concepts.

The first single layer graphene-based spin valves fabricated on a hBN substrate demonstrated 20~\textmu m distance spin transport, due to improved spin diffusion coefficient and high electronic mobility, see Fig.~\ref{fig:highQ}(a) \cite{zomer_long-distance_2012}. Surprisingly, spin relaxation times obtained in these devices exhibited comparable values to those obtained on conventional SiO$_2$ substrate. This observation was attributed to spin scattering due to fabrication-related residues, and {\red motivated pursuing} new studies including hBN encapsulation. 
Improved spin relaxation times up to 2~ns with spin relaxation lengths exceeding 12~\textmu m were observed in partially encapsulated graphene \cite{guimaraes_controlling_2014}. This device structure also allows studying the 
Rashba spin-orbit coupling and the resulting 
anisotropy of the spin relaxation time for spins pointing out-of-plane to spins pointing in-plane, and its modulation via perpendicular electric field, see Fig.~\ref{fig:highQ}(b). A similar geometry was also employed for bilayer spin valve devices, leading to spin relaxation lengths of 24~\textmu m \cite{ingla-aynes_24_2015}. 
\citet{avsar_electronic_2016} performed a comparative study by discussing the effect of the substrate and polymer residues on spin transport, where the observation of similar spin transport characteristics in non-encapsulated graphene on SiO$_2$ and hBN substrates suggested that spin transport in these devices was not limited by contacts, substrate phonons or impurities.

On the other hand, observation of a five-fold enhancement in spin relaxation times upon encapsulation, and a non-monotonic carrier concentration dependence of spin relaxation times, suggested that resonant scattering by unintentional magnetic impurities is the limiting source for spin scattering in graphene, consistent with a recent theoretical proposal \cite{kochan_resonant_2015}. 
These studies suggest that adapting a full encapsulation process could allow approaching graphene's intrinsic spin transport performance. Towards this, \citet{drogeler_spin_2016} developed a bottom-top approach for fabricating polymer free spin valves, where hBN encapsulated graphene was transferred on top of pre-fabricated Co/MgO spin electrodes. They observed record spin relaxation times of 12~ns with spin relaxation lengths exceeding 30~\textmu m, despite the presence of pin-holes in their MgO tunnel barriers, as shown in Fig.~\ref{fig:highQ}(c). These results also confirm the relevance of polymer residues on enhancing spin scattering events.

\subsubsection{Bilayer and few-layer graphene}

\begin{figure}
\includegraphics[width=1.0\columnwidth]{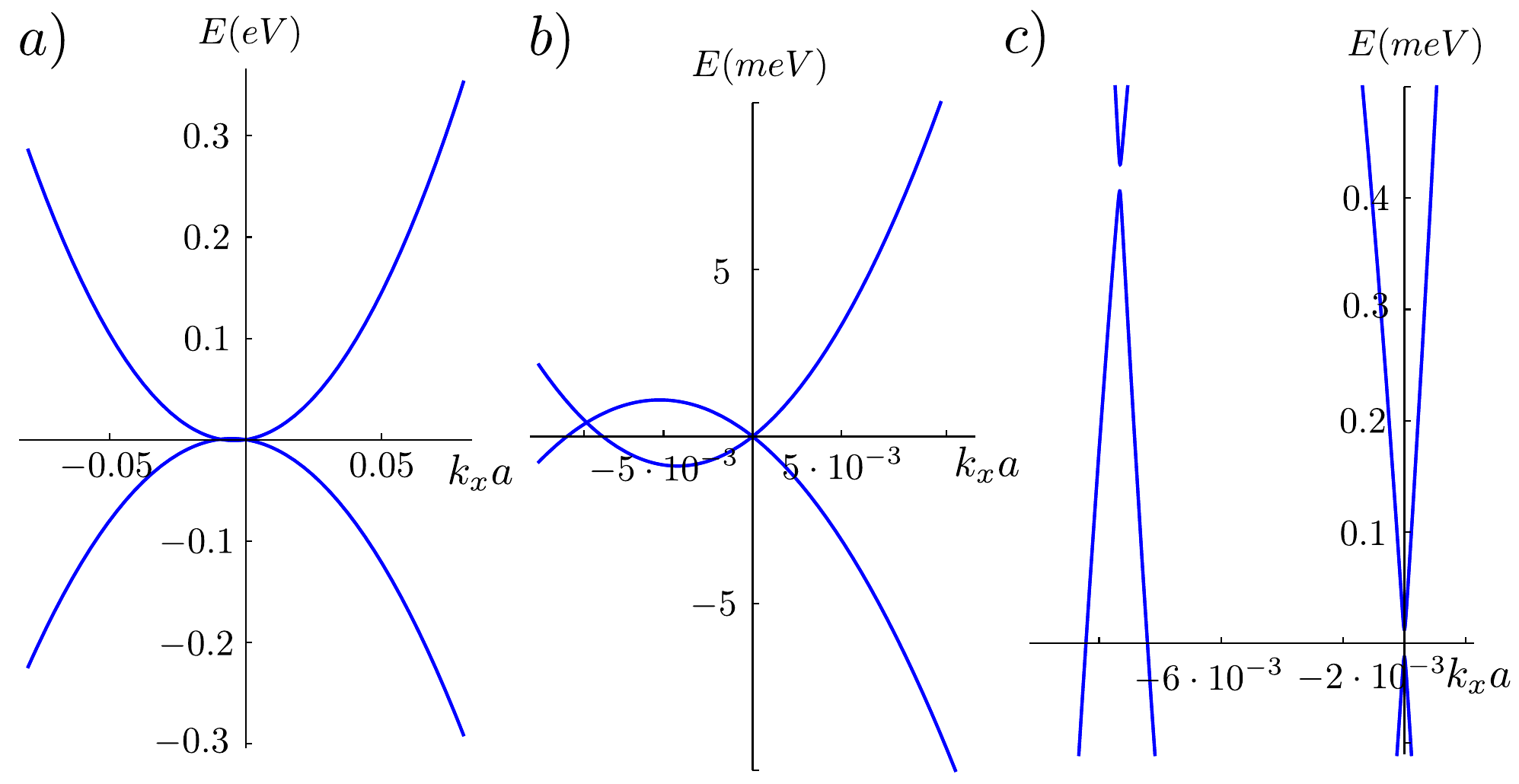}
\caption{
{\red Low-energy bands of bilayer graphene. a) Parabolic dispersion of the low-energy bands around one of the inequivalent corners of the Brillouin zone. The bands localized at the carbon atoms sitting on top of each other appear at higher energies. 
{\purple b) The two parabolas overlap due to the trigonal warping of the bands. 
The low energy bands were obtained using realistic DFT extracted tight-binding parameters.} 
c) The SOC terms $\Delta_{KM}=12$~\textmu eV and $\Delta_{BR}=19$~\textmu eV$\cdot$nm \cite{Konschuh_etal2} lift the band crossings. Note that the bands remain spin degenerate due to inversion symmetry.}
}
\label{fig:bands_bilayer}
\end{figure}

In bilayer graphene the unit cell contains 4 atoms. In the conventional Bernal stacking the low-energy bands are built up from $p_z$ orbitals localized at opposite sublattices in different layers. The bands touch at the corners of the Brillouin zone, as in the case of graphene, but with an approximately quadratic dispersion instead \cite{McCann_Falko1}.
This degeneracy is protected by the $D_{3d}$ point group symmetry of the crystal but it can be removed by applying a perpendicular electric field \cite{castro2007}. Since the bands are quadratic, we must consider spin-orbit couplings up to linear order in crystal momentum. Within the 2-bands effective model, the intrinsic terms compatible with the crystallographic symmetries read \cite{Guinea,Gelderen_Morais,McCann_Koshino,Konschuh_etal2} 
{\red
\begin{align}
\mathcal{H}_{so}^{\textrm{blg}}=\Delta_{KM}\sigma_z\tau_z s_z+\Delta_{BR}\left(k_x s_y-k_ys_x\right)\sigma_z.
\label{eq:bilayer}
\end{align}}
The first term is a Kane-Mele coupling, whereas the second adopts the form of the usual Bychkov-Rashba coupling in the 2D electron gas, 
{\red $k_x s_y-k_y s_x$}, 
but with opposite sign for electrons residing at different layers/sublattices. Notice that this term does not remove the spin degeneracy since the unit cell possesses a complete center of inversion ($D_{3d}=D_3\times i$, where $i$ is the inversion group). Another difference with respect to the case of single-layer graphene is that the Kane-Mele coupling is linear in the intra-atomic spin-orbit coupling due to the non-zero hybridization between $\pi$ and $\sigma$ orbitals localized at different layers. \textit{Ab initio} calculations give 
{\red $\Delta_{KM}=12$~\textmu eV, $\Delta_{BR}=19$~\textmu eV$\cdot$nm \cite{Konschuh_etal2}. The low-energy bands of bilayer graphene are shown in Fig.~\ref{fig:bands_bilayer}.}

This symmetry-based analysis can be extended to graphene multilayers with an arbitrary number of layers $N$ \cite{Guinea_etal1,Partoens_Peeters,Manes_etal}. The electronic properties of these systems depend both on $N$ and the type of stacking. A Bernal stack with even $N$ possesses $D_{3d}$ symmetry. The low-energy bands can be seen as $N/2$ copies of the low energy model of bilayer graphene. When $N$ is odd, however, the point group of the crystal is $D_{3h}$. In the simplest description, an additional band with linear (Dirac) dispersion appears, but none of these degeneracies is protected by the crystallographic symmetries in this case. The low-energy spectrum of rhombohedral stacks does not depend on the parity of the number of layers. There are only two bands that touch at corners of the Brillouin zone with dispersion 
{\red $|\mathbf{k}|^N$}. 
These bands become surface states localized at the top and bottom layers in the limit $N\rightarrow \infty$. The remaining $2N-2$ bands form Dirac crossings protected by $D_{3d}$ symmetry. Within the lowest-energy bands of centrosymmetric stacks only a Kane-Mele coupling is allowed by symmetry at 
{\red $\mathbf{k}=0$ }
(with the sublattice operators properly defined), whereas in the case of non-centrosymmetric stacks the spin-orbit coupling removes the spin degeneracy \cite{McCann_Koshino}. However, the spin polarization along the out-of-plane direction is still a good quantum number due to mirror symmetry ($D_{3h}=D_3\times\sigma_h$, where $\sigma_h$ is the mirror reflection along the basal plane).

\subsubsection{The original Topological insulator}

Despite its weakness, the spin-orbit coupling in graphene has attracted a lot of attention in the recent years due to the connection with the field of topological insulators \cite{Kane_Hasan,Qi_Zhang}. The intrinsic or Kane-Mele term removes the degeneracy at the Dirac points, opening a gap in the spectrum. Moreover, the virtual processes mediated by high energy bands resemble the pattern of complex hoppings proposed by Haldane as a model for the quantum anomalous Hall effect \cite{Haldane}. In this case, the phase $\pm \pi/2$ accumulated by the wave function is opposite for spin up and down. In a finite geometry, chiral modes localized at the edges of the system appear within the bulk gap, as shown in the calculation of Fig.~\ref{fig:qshe}. Backscattering between counter-propagating modes with opposite spin polarization is forbidden by time-reversal symmetry \cite{Kane_Mele2,Roy,Moore_Balents}. This quantum spin-Hall effect (QSHE) is unobservable in practice due to the narrowness of the topological gap. Nevertheless, it has been theorized that it could be stabilized by heavy adatom deposition \cite{weeks_engineering_2011}. 
{\purple Here, we note that such an adatom-induced topological gap has not been experimentally observed yet \cite{chandni_transport_2015, wang_electronic_2015, jia_transport_2015, wang_neutral-current_2015, santos_impact_2018}, since such dilute limit makes the realization of QSHE challenging \cite{milletari_crossover_2016}.}

The low SOC strength of graphene is detrimental for the implementation of graphene into many spintronics applications requiring high SOC strength such as spin FETs etc \cite{datta_electronic_1990}. Moreover, it also restricts the realization of theoretically predicted quantum spin Hall state at experimentally accessible temperatures \cite{Kane_Mele1}. All these stimulate the development of new methods for extrinsically enhancing SOC. 

\subsection{Corrugations and resonant impurities} \label{sec:impurities}

Corrugations naturally break the mirror symmetry of graphene, 
{\red giving rise to new SOC terms. Microscopically, the origin of these new couplings is the change of relative orientation of the orbitals due to the extrinsic curvature (i.e., the bending) of the graphene sheet, which hybridizes the $\pi$ and $\sigma$ bands (otherwise, hopping between these orbitals would be precluded by mirror symmetry).}
In a low-energy continuum description, this extrinsic curvature is characterized by a second rank tensor, $\mathcal{F}_{ij}\approx\partial_i\partial_j h$ with $i,j=x,y$, describing the embedding of the graphene sheet in ambient space, where the field $h(x,y)$ represents the out-of-plane displacement of carbon atoms at position $(x,y)$. The mean curvature $\mathcal{F}_{ii}\approx\nabla^2 h$ generates a Bychkov-Rashba coupling of the form \cite{Huertas_etal,Jeong_etal,Ochoa_etal1},
\begin{align}
\mathcal{H}_{BR}=g_{BR}\,\mathcal{F}_{ii}\left(\tau_z\sigma_x s_y-\sigma_y s_x\right).
\label{eq:bent1}
\end{align}
The tensorial nature of $\mathcal{F}_{ij}$ allows for additional couplings, incorporating the effect of a preferential direction of bending \cite{Ochoa_etal1}:
\begin{align}
\mathcal{H}_{BR}'=\, & g_1\left[\left(\mathcal{F}_{xx}-\mathcal{F}_{yy}\right)\tau_zs_y+2\mathcal{F}_{xy}\tau_zs_x\right]\nonumber\\
&+ g_2[\left(\mathcal{F}_{xx}-\mathcal{F}_{yy}\right)\left(\tau_z\sigma_x s_y+\sigma_y s_x\right)\nonumber\\
& +2\mathcal{F}_{xy}\left(\sigma_y s_y-\tau_z\sigma_x s_x\right)].
\label{eq:bent2}
\end{align}
From the previous tight-binding estimates we have \cite{Jeong_etal,Ochoa_etal1}\begin{align}
& g_{BR}=\frac{a\epsilon_s\Delta_{SO}\left(V_{pp\pi}+V_{pp\sigma}\right)
}{12V_{sp\sigma}^2}\approx 1.2  \text{ meV}\cdot\text{\AA},\nonumber\\
& g_1=\frac{aV_{pp\pi}\Delta_{SO}}{2\left(V_{pp\sigma}-V_{pp\pi}\right)}\approx1.5\mbox{ meV}\cdot\text{\AA}.
\end{align}
where $V_{pp\pi}\approx-2.2$ eV and $V_{pp\sigma}\approx5.4$ eV represent the hopping between $p$-orbitals at nearest neighbors. The second coupling in Eq.~\eqref{eq:bent2} only appears at second nearest neighbors, so it is usually neglected.


Since graphene is all surface, its environment sensitivity enables different ways of boosting its SOC including functionalization, adatom decoration and substrate engineering \cite{castro_neto_impurity-induced_2009, weeks_engineering_2011, ma_strong_2012, irmer_spin-orbit_2015, gmitra_spin-orbit_2013, soriano_spin_2015, pachoud_scattering_2014, ferreira_extrinsic_2014}. 
One of the first proposed strategies to enhance the SOC was the functionalization of graphene surface with hydrogen adatoms, which create a local out-of-plane, $sp^3$-like distortion of the surrounding bonds. 
This distortion hybridizes the $\sigma$ states with the adjacent $p_z$ orbitals, giving rise to an enhancement of the spin-orbit coupling \cite{castro_neto_impurity-induced_2009},\begin{align}
\Delta\approx\tan\vartheta\,\sqrt{1-2\tan^2\vartheta}\,\Delta_{SO},
\end{align}
where $\vartheta$ represents the angle of the distorted $\sigma$-bond with respect to the graphene plane (i.e., $\vartheta=0$ for the flat configuration). 
In the case of a complete $sp^3$ hybridization ($\vartheta=19.5^{\circ}$) we have an enhanced SOC up to $\Delta\approx 6$ meV. 
The symmetry is effectively reduced down to $C_{3v}$, so the specific couplings in the low-energy theory acquire different forms \cite{gmitra_spin-orbit_2013}. \textit{Ab initio} calculations give $\Delta_{BR}\approx0.33$ meV for the Bychkov-Rashba coupling. Notice that in this situation the bottleneck is still the weak intra-atomic spin-orbit interaction in carbon. 

\begin{figure*}
\includegraphics[width=1.0\textwidth]{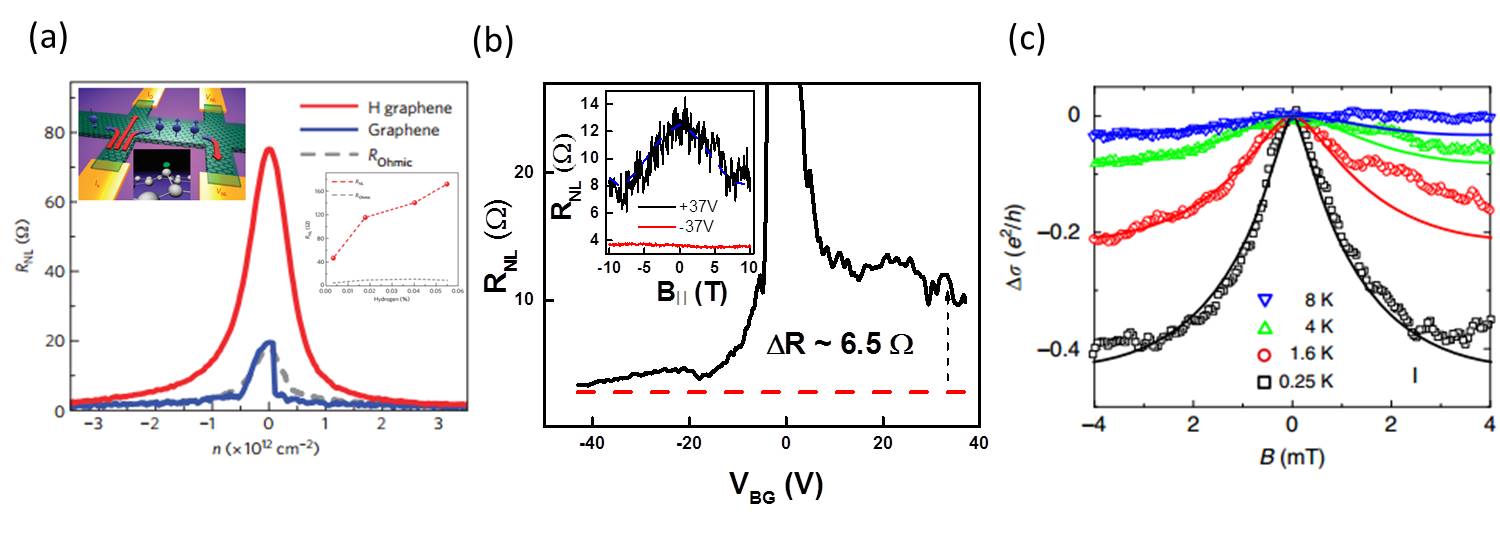}
\caption{
\emph{Exploiting Spin-Orbit Coupling in Graphene.}
(a) Carrier concentration dependence of the nonlocal resistance in weakly hydrogenated and pristine graphene, shown as red and blue lines, respectively. The dotted gray line represents the calculated Ohmic contribution. 
The inset shows the dependence of the nonlocal resistance on the hydrogenation percentage. From \citet{balakrishnan_colossal_2013}. 
(b) Back gate voltage ($V_\text{BG}$) dependence of the nonlocal resistance in graphene devices supported on WS$_2$. Spin precession and a sizeable signal are only detectable for $V_\text{BG}>0$. From \citet{avsar_spinorbit_2014}. 
(c) Magneto-transport measurements for graphene on WS$_2$ substrate, showing a weak anti-localization effect. 
From \citet{wang_strong_2015}.
}
\label{fig:exploting-soc}
\end{figure*}

\citet{balakrishnan_colossal_2013} experimentally reported a SOC enhancement of 2.5~meV in weakly hydrogenated graphene samples, see Fig.~\ref{fig:exploting-soc}(a). Such enhancement caused the observation of spin Hall effect (SHE) at room temperature, with the spin-transport origin of the nonlocal signal confirmed by spin precession measurements. Similar results were also obtained in fluorinated graphene samples \cite{avsar_enhanced_2015}. 
Nevertheless, other studies have reported large nonlocal signals in similarly hydrogenated devices but with an absence of any magnetic field dependence in spin precession measurements \cite{kaverzin_electron_2015}, 
{\red or contradicting the expectation of a decrease in spin lifetime due to larger SOC \cite{wojtaszek_enhancement_2013}.} 
These latest results suggest that the effect is sample dependent and there can be additional mechanisms involved other than spin transport, as theoretically suggested for the case of disorder in graphene \cite{van_tuan_spin_2016}. 
{\red These conflicting interpretations for experiments in hydrogenated graphene call for further studies to ensure a thorough verification of the role of SOC and the presence of the SHE in this system, and to further identify any other concomitant effect leading to the nonlocal response observed.}

The passivation of $p_z$ orbital below the adatom creates a \textit{perfect} vacancy (i.e., with no reconstruction altering the coordination of the lattice). Such a defect induces the formation of a quasi-localized state, manifested as a sharp resonance in the local density of states \cite{Pereira_etal,Wehling_etal}. The enhancement of the density of states at energies close to the Dirac point favors the formation of magnetic moments due to electron-electron interactions \cite{Palacios_etal,Yazyev_2008}, which can modify both charge and spin transport properties. Non-perturbative calculations beyond mean field show no signature of saturation of the susceptibility at low temperatures \cite{Haase_etal}, suggesting a ferromagnetic coupling between the quasi-localized magnetic moment and the itinerant spins. Nevertheless, experiments have so far demonstrated a paramagnetic response \cite{nair_spin-half_2012, mccreary_magnetic_2012}.


Another direction to enhance the weak SOC of graphene is the decoration of its surface with heavy metallic adatoms \cite{weeks_engineering_2011, pachoud_scattering_2014, ferreira_extrinsic_2014, Brey_2015, hu_giant_2012, ma_strong_2012, Cysne2018}. In this approach, graphene's $sp^2$ bond property is preserved; the SOC is locally enhanced due to tunneling of electrons from graphene to adatoms and back. \citet{marchenko_giant_2012} observed a giant spin orbit splitting of $\approx$ 0.1~eV in Au intercalated graphene samples. The photoelectron spectroscopy measurements revealed that hybridization with Au $5d$ states is the source of spin orbit splitting. 
Large spin orbit fields were also achieved in Pb intercalated graphene on Ir substrate \cite{Calleja_etal}. \citet{balakrishnan_giant_2014} reported spin Hall effect at room temperature in CVD grown graphene devices, 
attributed to unavoidable residual Cu adatom clusters, 
with a SOC of $\approx$ 20 meV. 
However there are experimental controversies over nonlocal measurements in adatom decorated samples as well. For example, other groups also reported large nonlocal signals in Au and Ir decorated samples, but they did not observe any magnetic field dependence \cite{wang_neutral-current_2015}. Further work is therefore needed to address the possible role of valley currents or of variations in adatom cluster sizes.

\subsection{Transition-metal dichalcogenides and graphene}

The family of atomically thin 2D crystals goes beyond the allotropes of carbon and already includes materials like phosphorene \cite{Xia_etal}, graphane C$_2$H$_2$ \cite{graphane}, or monolayers of hexagonal boron nitride (hBN) \cite{hBN}, among others \cite{geim_grigorieva}. Silicene \cite{Vogt_etal, Fleurence_etal}, a single layer of silicon atoms forming a $sp^3$-like honeycomb lattice, has attracted much attention due to its resemblance with graphene. The spin-orbit coupling within the $\pi$ bands reduces to Eq.~\eqref{eq:bilayer} as imposed by $D_{3d}$ point-group symmetry. Interestingly, the Kane-Mele coupling is much larger than in graphene, $\Delta_{KM}\sim1.5$ meV, due to its buckled structure. For the same reason, the band topology can be controlled by applying a perpendicular electric field \cite{Drummond_etal,Ezawa}.

The most appealing materials regarding spintronics applications are probably monolayers of transition-metal dichalcogenides (TMDC) \cite{MX2_review}. Bulk compounds are composed of X-M-X layers (X representing the chalcogen atoms, M the transition metal) stacked on top of each other and coupled by weak van der Waals forces. They show different polytypes which vary in stacking and atom coordination \cite{MX2_review}. Like graphite, these materials can be exfoliated down to a single layer \cite{MX2}. Semiconducting materials include molybdenum disulfide (MoS$_2$), tungsten disulfide (WS$_2$), molybdenum diselenide (MoSe$_2$), or tungsten diselenide (WSe$_2$) \cite{Mak_etal_prl}.


The point group of the monolayer crystal is $D_{3h}$. The lattice consists in a triangular Bravais lattice with two X atoms and one M atom per unit cell. As in the case of graphene, the Fermi level lies around the two inequivalent corners of the hexagonal Brillouin zone. However, the dispersion and orbital character of the conduction and valence bands are completely different. The large crystal fields associated with the different atomic species prevent accidental degeneracies, so the bands remain gapped with approximately quadratic dispersion \cite{Cappelluti_etal,Rostami_etal}. The spin-orbit interaction splits the spin degeneracy of the bands,\begin{align}
\mathcal{H}_{so}^{\textrm{tmd}}=\lambda_c\,\frac{1+\sigma_z}{2}\tau_z s_z+\lambda_v\,\frac{1-\sigma_z}{2}\tau_z s_z.
\end{align}
The magnitudes of the splittings in conduction and valence bands are very different due to their distinct orbital character, dominated by $d$ orbitals from the transition metal in both cases. A complete tight-binding model can be found in~\onlinecite{Roldan_etal}. Values extracted from \textit{ab initio} calculations \cite{Zhu_etal,Feng_etal,Kormanyos_etal} are summarized in Table~\ref{tab:splittings}. Notice that the splitting of the bands has opposite sign at each valley, as imposed by time-reversal symmetry. This spin-valley coupling \cite{Xiao_etal} enables the optical control of valley populations \cite{Cao_etal,Mak_etal,Zeng_etal}. As will be discussed in Section \ref{sec:opto}, this property allows generation of spin polarized charge carriers, without the need for a ferromagnetic spin injector.

\begin{table}
\begin{tabular}{|c|c|}
\hline
Material&$\lambda$ (meV)\\
\hline
\hline
MoS$_2$ (conduction band) &3\\
\hline
MoS$_2$ (valence band) &140\\
\hline
WS$_2$ (valence band) & 430\\
\hline
MoSe$_2$ (valence band) &180\\
\hline
WSe$_2$ (valence band) & 460\\
\hline
\end{tabular}
\caption{Band spin splittings in monolayers of transition-metal dichalcogenides. Values from  \cite{Zhu_etal,Feng_etal,Kormanyos_etal}.}
\label{tab:splittings}
\end{table}



\begin{figure}
\includegraphics[width=1.0\columnwidth]{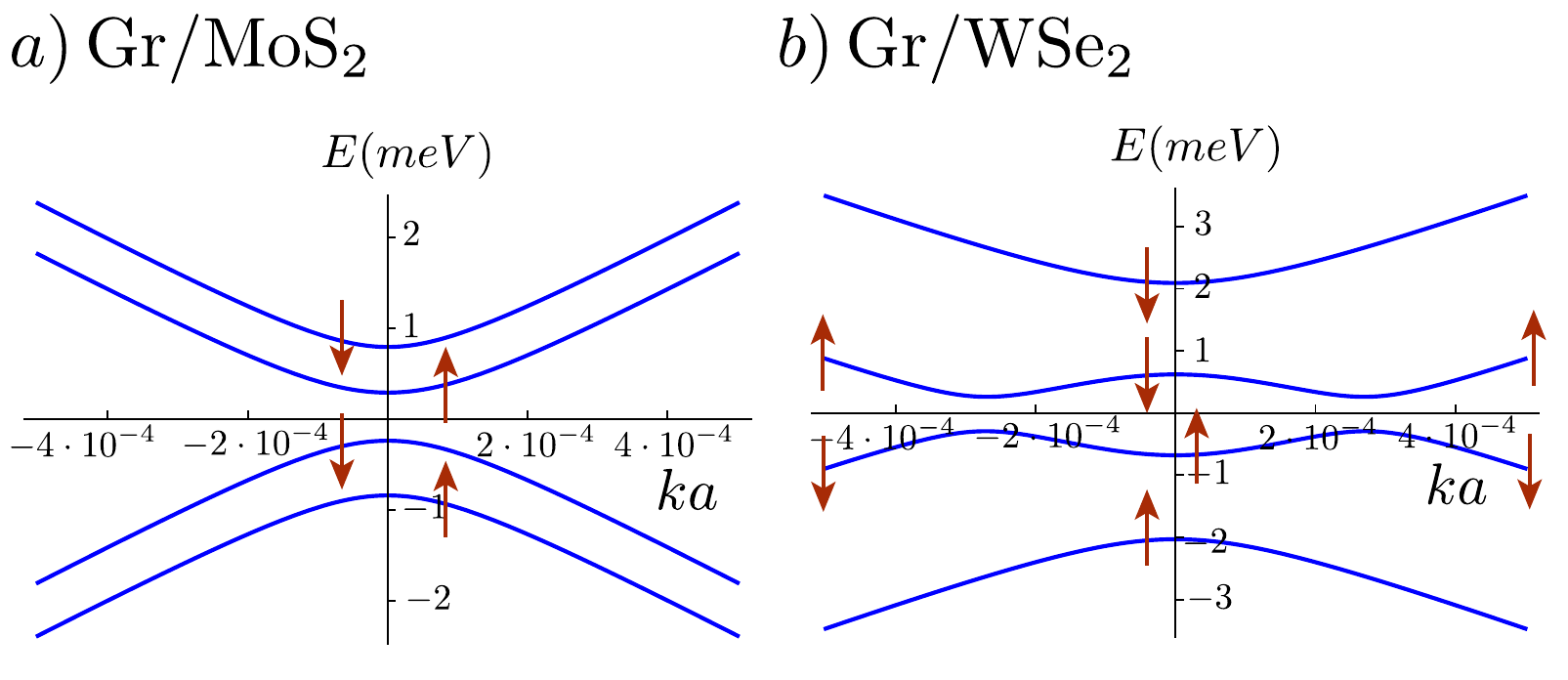}
\caption{Low-energy bands of graphene on MoS$_2$ (a) and WSe$_2$ (b) deduced from the Hamiltonian in Eq.~\eqref{eq:graphene/TMDC}. The values of the parameters correspond to \cite{Gmitra_etal_2016}: a) $\Delta_{st}=0.52$, $\Delta_{KM}=-0.025$, $\Delta_{BR}=0.13$, $\lambda=-0.255$, and b) $\Delta_{st}=0.54$, $\Delta_{KM}=-0.03$, $\Delta_{BR}=0.56$, $\lambda=-1.19$ (all the values in meVs). The arrows indicate the (approximate) spin polarization of the bands, opposite at different valleys.}
\label{fig:heterostructures}
\end{figure}

Graphene placed on top of a TMDC forms a very appealing heterostructure from the point of view of spintronics. The Dirac cone lies within the gap of the transition-metal dichalcogenide, so the low-energy $\pi$-bands preserve their identity while acquiring a remarkably large spin-orbit coupling by proximity with the heavy atoms in the second layer \cite{Kalomi_etal,Gmitra_etal_2016,Wang_etalPRX}. The dispersion is well described by \cite{gmitra_graphene_2015,Alsharari_etal,Kochan_etal_2017}
\begin{align}
\label{eq:graphene/TMDC}
& \mathcal{H}_{\textrm{gr/tmd}}=\hbar v_F\left(\tau_z\sigma_x k_x+\sigma_y k_y\right)+\Delta_{st}\sigma_z\\
& + \Delta_{KM}\tau_z\sigma_z s_z +\Delta_{BR}\left(\tau_z\sigma_x s_y-\sigma_y s_x\right)+\lambda \tau_z s_z.\nonumber
\end{align}
Compared to Eq.~\eqref{eq:Dirac}, two new terms appear in this Hamiltonian: the staggered potential $\Delta_{st}$, which opens a trivial gap in the spectrum, and the last term in the second line of Eq.~\eqref{eq:graphene/TMDC}, which removes the spin degeneracy of the bands. Both terms reflect that the symmetry of graphene is effectively reduced down to $C_{3v}$ due to the interaction with the underlying TMDC system. The model can be generalized to describe heterostructures with bilayer graphene, for which the induced spin-orbit coupling can be tuned by applying a perpendicular electric field \cite{Khoo_etal,Gmitra_Fabian_2017}. 
{\purple Such tunability makes bilayer graphene on WSe$_2$ an efficient field-effect spin-orbit valve \cite{Gmitra_Fabian_2017}.} 

Two distinctive band dispersions arise from this model, as represented in Fig.~\ref{fig:heterostructures}. In the first case (a) the staggered potential dominates over the induced spin-orbit coupling, whereas in the second (b) the spin-orbit coupling produces a band inversion. This second case is expected to occur for the heaviest atomic species in the second layer (namely, W and Se). The band inversion resembles a time-reversal symmetric version of the quantum anomalous Hall effect proposed in graphene interacting with a heavy magnet \cite{Qiao_etal_I,Qiao_etal_II}, where the term parametrized by $\lambda$ can be interpreted as a valley-dependent exchange coupling. However, this band inversion does not correspond to a topological state, like in the idealized case of pristine graphene (Fig.~\ref{fig:qshe}): in general, two (instead of one) pairs of sub-gap helical modes connected by time-reversal symmetry appear at the boundaries of the system \cite{Yang_etal2D}. Backscattering between states belonging to different Kramers doublets (in a zig-zag ribbon, one pair around the projected $K$ and $K'$ points and the other at $M$) is only precluded by crystalline symmetries, which are removed by generic disorder. Moreover, the edges states are completely absent in armchair ribbons. 

The approaches for enhancing SOC in graphene previously discussed in Section \ref{sec:impurities}, regarding functionalization \cite{avsar_enhanced_2015, balakrishnan_colossal_2013} and adatom decoration \cite{balakrishnan_giant_2014}, had the downside of reducing the electronic quality of graphene. On the other hand, creating a van der Waals interface between graphene and a TMDC can enhance the SOC of the former while simultaneously preserving its high electronic mobility \cite{tan_avsar_2014}. Towards this, \citet{avsar_spinorbit_2014} reported a three orders of magnitude enhancement of SOC strength by bringing graphene into proximity with WS$_2$. The resulting large SOC strength of up to 17.6~meV led to the observation of (I)SHE for transport in the electron regime, attributed to sulfur-based in-gap defect states, see Fig.~\ref{fig:exploting-soc}(b). The large spin orbit coupling in sulfur vacancies in monolayer WS$_2$ has been recently confirmed by detailed scanning probe microscopy studies \cite{schuler_large_2018}. 

The relative strength of the induced spin-orbit couplings can be extracted from magneto-transport experiments. 
\citet{wang_strong_2015} demonstrated pure interface-induced SOC enhancement by performing weak anti-localization measurements, see Fig.~\ref{fig:exploting-soc}(c). 
The last two terms in Eq.~\eqref{eq:graphene/TMDC} possesses different symmetry with respect to mirror reflection, which provides different signatures in the quantum-interference correction to the conductivity. In the absence of inter-valley scattering, the magneto-conductance is given by \cite{McCann_Falko2}
\begin{align}
\label{eq:magnetoconductance}
 \Delta G=\, & -\frac{e^2}{2\pi h}\left[F\left(\frac{B}{B_{\phi}}\right)-F\left(\frac{B}{B_{\phi}+2B_{\textrm{asy}}}\right)\right.\nonumber\\
& \left.-2F\left(\frac{B}{B_{\phi}+B_{\textrm{sym}}+B_{\textrm{asy}}}\right)\right],
\end{align}
where $F(z)=\log z+\Psi(1/2+1/z)$, $\Psi(x)$ is the digamma function, and $B_{i}\equiv \hbar/(4e\ell_i^2)$. Here $\ell_{\textrm{sym}(\textrm{asy})}$ correspond to the spin diffusion lengths limited by mirror symmetric (asymmetric) spin-orbit terms, whereas $\ell_{\phi}$ represents the phase-coherence length limited by inelastic scattering (phonons or electron-electron interactions). While spin dephasing induced by mirror asymmetric terms (i.e., the Bychkov-Rashba coupling) should be manifested as a weak anti-localization effect at low temperatures and fields, the mirror symmetric terms (i.e., the spin-valley locking term $\lambda$) introduces an effective saturation of the decoherence times, leading to a suppression of the magneto-conductance at the lowest fields ($B<B_{\textrm{sym}}$). A similar crossover is expected in TMDC monolayers \cite{Ochoa_etal4}. 
In hybrid graphene-TMDC systems, a weak anti-localization behavior is systematically reported, as discussed previously, in a wide range of carrier concentrations \cite{wang_strong_2015, Wang_etalPRX, Yang_etal_PRB, Yang_etal2D}, suggesting a dominance of mirror asymmetric terms. 
Extracting the strength of the couplings requires the knowledge of the underlying spin-relaxation mechanism, which is the subject of Section \ref{sec:relaxation}. 
{\purple Alternatively, for quantizing fields one can trace the effect of different SOC couplings in the sequence of Landau levels \cite{Cysne2018b, Khoo2018}.} 
Overall, the observation of similar results with different dichalcogenides indicates the robust and strong nature of proximity-induced SOC enhancement in graphene-TMDC heterostructures. 

{\purple The possible topological phases created by proximity-induced SOC are largely enriched in graphene multilayers. In that case, the band topologies crucially depend on the (parity) number of layers and stacking order, and can be tuned by displacement field \cite{Zaletel2019}. The fundamental observation is that layer-resolved spin-valley couplings enter as an effective Kane-Mele mass in the low-energy chiral bands of rhombohedral stacks due to the equivalence between layer and sublattice projections. In bilayer graphene/TMDC heterostructures, recent electronic compressibility measurements have revealed a SOC-induced band inversion compatible with $\lambda=1.7$ meV \cite{Island2019}, very close to theoretical estimations. In these devices, magnetotransport data suggests the existence of helical edge channels. Although backscattering (similarly to the monolayer case discussed above) is not forbidden by time-reversal symmetry, these states are expected to be more resilient against mirror asymmetric (Rashba-like) couplings thanks to interlayer coupling.}

\begin{figure*}
\includegraphics[width=1.0\textwidth]{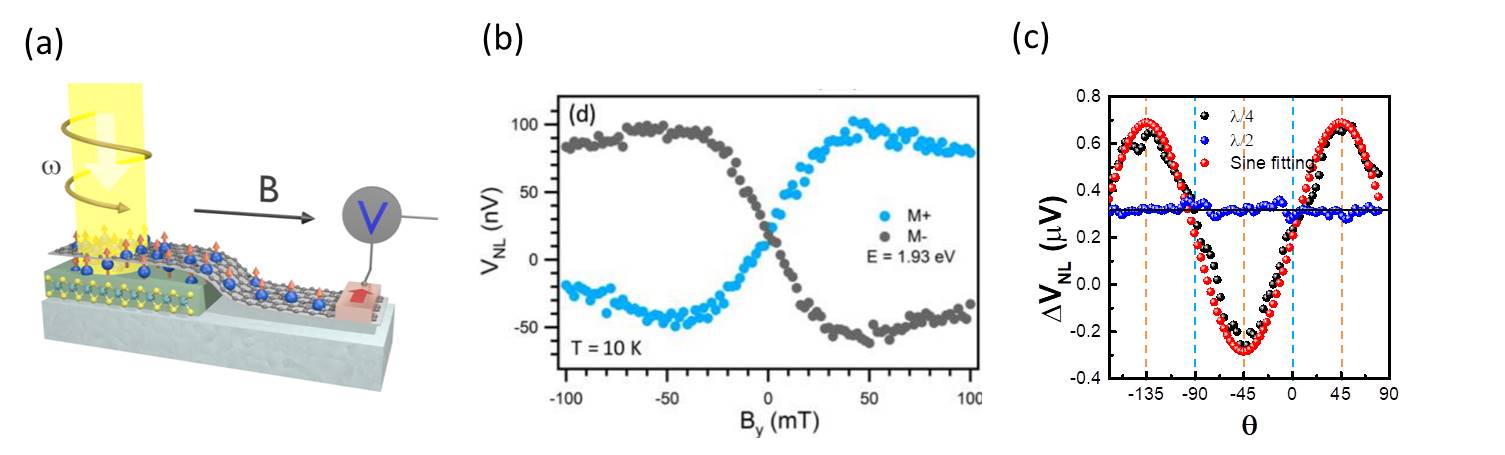}
\caption{
\emph{Alternative Spintronics in Graphene and Other 2D Materials.}
(a) Schematics of optical spin injection in graphene (gray) by utilizing a monolayer transition metal dichalcogenide (green) and excitation with circularly polarized light. From \citet{gmitra_graphene_2015}. 
(b) Hanle spin precession curves for different detector magnetization directions in a graphene/MoS$_2$ device. From \citet{luo_opto-valleytronic_2017}. 
(c) Incident angle dependence of optically generated nonlocal signal under quarter wave modulation in  a graphene/WSe$_2$ device. From \citet{avsar_optospintronics_2017}.
}
\label{fig:alternative}
\end{figure*}

\subsection{Graphene/TMDC optospintronics} \label{sec:opto}

Non-destructive optical spin injection into graphene has the potential to overcome limitations inherent to electrical spin injection methods whilst enabling new opto-spintronic functionalities. 
The weak optical absorption and weak intrinsic SOC strength of graphene prevents such a direct optical spin injection \cite{inglot_optical_2014}. 
{\purple Silicon also poses a similar challenge towards spin injection, due to its low SOC strength. Combining Si with a direct band gap semiconductor has been proposed to overcome this challenge \cite{zutic_spin_2006}.} 
In this regard, \citet{gmitra_graphene_2015} proposed the use of proximity-induced SOC as way to enable direct optical spin injection. Here, graphene is partially covered by a monolayer transition-metal dichalcogenide crystal, as shown in Fig.~\ref{fig:alternative}(a). Excitation with circularly polarized light on the TMDC-covered graphene area generates spin polarized charge carriers in the TMDC monolayer, as a result of spin-valley coupling and the valley-selective absorption. These carriers diffuse into the adjacent graphene layer, and are finally detected electrically using ferromagnetic contacts in a non-local geometry. 
This non-destructive optical spin injection schemes were recently experimentally realized independently by two groups. \citet{luo_opto-valleytronic_2017} utilized monolayer MoS$_2$ to inject spin current into multilayer graphene at room temperature. They utilized Hanle spin precession measurements to confirm optical spin injection, spin transport and electrical detection, see Fig.~\ref{fig:alternative}(b). \citet{avsar_optospintronics_2017}, utilized monolayer WSe$_2$ to inject spin current into monolayer graphene, where the generated nonlocal spin signal was electrically detected by utilizing hBN as a tunnel barrier. The spin-coupled valley selective origin of the signal was confirmed by prudently studying the dependences of the nonlocal spin signal on modulations of the incident light intensity and polarization, see Fig.~\ref{fig:alternative}(c). In future experiments it will be relevant to determine the types of charge carriers diffusing into graphene, as well as experimentally demonstrating the optical detection of such generated spin signal, free from any contact-related spin scattering at FM/NM interfaces.

\section{Current challenges and way forward}

\subsection{What is the dominant spin relaxation mechanism?} \label{sec:relaxation}

The polarization of a spin current in a metal or semiconductor device decays during the propagation of the spin carriers. Defining a characteristic time of decay in a rigorous way is a difficult task, though. This is usually introduced within the framework of the Bloch-Torrey equations \cite{Bloch,Torrey} describing the macroscopic dynamics of the spin density in the material. 
Related phenomenological models can be used to fit the experimental data and extract characteristic times of spin relaxation. However, a theoretical evaluation requires more sophisticated models for the microscopic spin dynamics. Once this is determined, a quantum kinetic equation for the spin-ensemble dynamics can be derived \cite{Mishchenko_etal}. Microscopic expressions for the relaxation rates can be obtained in the limit of long wave-lengths, in which the variations of the density matrix are smooth on the scale of the characteristic length of the problem, e.g., the mean-free path $\ell$ in the diffusive regime \cite{Burkov_etal}. This procedure is usually simplified by a semi-classical treatment based on the spin-Boltzmann equation. This approach has been applied to the study of spin relaxation in single-layer \cite{Zhou_Wu,Dugaev_etal,Zhang_Wu_2011,Zhang_Wu_2012} and bilayer \cite{Diez_Burkard} graphene, and more recently in transition-metal dichalcogenides \cite{Wang_Wu1,Wang_Wu2}. 

Four main mechanisms of spin relaxation are usually discussed for metals and semiconductors \cite{zutic_spintronics:_2004}: the Elliot-Yafet \cite{Elliot,Yafet}, D'yakonov-Perel' \cite{DyakonovPerel,Dyakonov}, Bir-Aronov-Pikus \cite{Bir_etal_1976}, and hyperfine-interaction mechanisms \cite{Dyakonov_Perel_1973}. The last two are usually disregarded in the context of graphene. On the one hand, the natural abundance of carbon isotopes with nuclear magnetic moment is very low; also, the nuclear fields acting on the spin of itinerant electrons averages-out in the diffusive limit, so this contribution can be safely neglected in the experimental situations that we are going to discuss here \cite{wojtaszek_absence_2014}. On the other hand, the Bir-Aronov-Pikus mechanism accounts for electron spin-flip processes mediated by the electron-hole exchange interaction, and therefore it is only relevant in heavily $p$-doped semiconductors.

The Elliot-Yafet (EY) mechanism takes into account the change in the spin polarization of a Bloch electron due to scattering off phonons or impurities. It is characterized by a linear relation between the spin and the quasi-momentum relaxation times, $\tau_s=\alpha\,\tau_p$, where $\alpha$ can be interpreted as the spin-flip probability during a scattering event. This relation can be deduced by treating the spin-orbit interaction perturbatively and holds experimentally for most conventional metals (i.e., systems with well-defined Fermi surface). In doped graphene, the EY mechanism is dominated by inter-band transitions and the Elliot relation depends explicitly on the carrier concentration \cite{HuertasHernandoetalprl,Ochoa_etal2},\begin{align}
\label{eq:EY}
\frac{1}{\tau_s}\approx\left(\frac{\Delta}{\epsilon_F}\right)^2\tau_p^{-1}.
\end{align}
Here $\Delta$ represents the strength of the spin-orbit interaction within the $\pi$-bands and $\epsilon_F=\hbar v_F \sqrt{\pi n}$ is the Fermi energy with respect to the Dirac point, where $n$ represents the carrier concentration. For typical values of these parameters, $\Delta\approx10$~\textmu eV, $n=10^{12}$ cm$^{-2}$, and $\tau_p=10$ fs, the spin relaxation times due to the EY mechanism are of the order of $\tau_s\approx10$~\textmu s, $\approx 4$ orders of magnitude longer than the relaxation times reported in Hanle precession experiments. 

The D'yakonov-Perel' (DP) mechanism accounts for the spin dephasing in between scattering events. As we discussed earlier, the doubly degeneracy of the bands is lifted in non-centrosymmetric crystals, and the spin-orbit coupling can be interpreted as an effective Zeeman field that makes the electron spin to precess in the Bloch sphere. The axis of precession depends on the direction of motion of the electron, and therefore scattering randomizes the process. This is an example of motional narrowing, characterized by inverse relation between the spin and the quasi-momentum relaxation times \cite{Fabian_rev},\begin{align}
\label{eq:DP}
\frac{1}{\tau_s}\approx \left(\frac{\Delta_{BR}}{\hbar}\right)^2 \tau_p.
\end{align}
The term in parentheses must be interpreted as the effective Larmor frequency associated with a Bychkov-Rashba coupling generated by, e.g.\ the interaction with the substrate \cite{Ertler_etal}. Realistic estimates lead to $\tau_s\approx 1$~\textmu s, again much longer than the relaxation times reported in Hanle precession. This discrepancy in several orders of magnitude along with the shortcomings of Eqs.~\eqref{eq:EY}-\eqref{eq:DP} to reproduce the intricate behavior of $\tau_s$ as a function of doping or temperature (particularly in the case of bilayer graphene) has provoked a tremendous activity in the recent years.

\subsubsection{Intrinsic relaxation sources}

As we discussed earlier, out-of-plane displacements enhance the strength of the spin-orbit coupling within the $\pi$-bands due to hybridizations with higher $\sigma$-states. Therefore, flexural phonons (the quanta of these vibrations) represent an unavoidable source of spin decoherence in clean, free-standing graphene. In fact, flexural phonons constitute the leading scattering mechanism in suspended samples at temperatures $T\gtrsim10$ K, limiting the electron mobilities down to few m$^2$/Vs \cite{castro2010}. Their contribution to spin relaxation can be evaluated from the couplings in Eqs.~\eqref{eq:bent1}-\eqref{eq:bent2}, leading to \cite{Fratini_etal,Vicent_etal}\begin{align}
\frac{1}{\tau_s}=\frac{\tilde{g}^2 \,k_F}{2\hbar^2 v_F}\,\frac{k_B T}{\kappa}\left(\frac{2k_F}{q_c}\right)^{2\nu},
\label{eq:tau_s-phonons}
\end{align}
where $\tilde{g}^2\equiv g_{BR}^2+g_1^2/4+g_2^2$ is the effective spin-phonon coupling. Here $\nu=0,1$ corresponds to the free-standing regime ($q_c\ll 2k_F$, where $k_F$ is the Fermi momentum of carriers and $q_c$ represents an infrared cut-off of the harmonic phonon dispersion) and samples under tension, respectively. In the former case, the dispersion relation of flexural phonons is approximately quadratic, $\omega_{\mathbf{q}}=\sqrt{\kappa/\rho}|\mathbf{q}|^2$, where $\kappa\approx 1$ eV \cite{kudin_etal} is the bending rigidity. At momenta $|\mathbf{q}|<q_c$ the dispersion relation is effectively linearized due to anharmonic effects \cite{Zakharchenko_etal} or applied tensions, altering the dependence on the Fermi wavevector $k_F=\sqrt{\pi n}$ when the strain exceeds $\bar{u}\approx 4\pi n \kappa /K$, $K\approx21$ eV\AA$^{-2}$ being the two-dimensional bulk modulus \cite{Lee_etal_2008}. In free-standing graphene, the spin relaxations times are of the order of hundreds of nanoseconds at room temperature, rapidly decreasing when tension is applied. Note also that the spin-relaxation rate in Eq.~\eqref{eq:tau_s-phonons} grows linearly with the temperature, in contrast to momentum scattering, dominated by two-phonon processes. This leads to a distinctive $T^2$-dependence in the mobilities of suspended samples \cite{castro2010}, so the spin-relaxation times are expected to deviate from Eq.~\eqref{eq:EY} in the clean limit.

In supported samples, graphene remains pinned to the substrate and the contribution from flexural phonons is strongly suppressed. Nevertheless, the interaction with the substrate enhances the spin-orbit coupling and induces spin relaxation \cite{Ertler_etal}. On top of that, supported graphene sheets also present static corrugations with characteristic heights of $h_0\approx 0.3$ nm \cite{Locatelli_etal}, which in the simplest approximation creates a non-uniform Bychkov-Rashba coupling. In the diffusive limit, we end up with two different regimes of spin relaxation arising from the competition between the two relevant length scales in the problem, the mean free path, $\ell$, and the lateral size of the ripples, $\mathcal{L}$. In the scattering-dominated regime, $\ell\ll\mathcal{L}$, spin decoherence is limited by the conventional motional narrowing. In between scattering events, the electron spin precesses an angle $\delta$, which can be estimated from the Larmor frequency $\Delta_{BR}/\hbar$, where $\Delta_{BR}\approx g_{BR}\times h_0/\mathcal{L}^2$ is the average strength of the corrugation-induced spin-orbit coupling. 
After $N$ collisions, and assuming that the process is Markovian, the precession angle is about $\delta(N)\sim\Delta_{BR}/\hbar\times\tau_{p}\times\sqrt{N}$. The spin-relaxation time may be defined as the time after which the precession angle is $\delta(N_s=\tau_s/\tau_{p})\sim 1$, from which we deduce the DP relation in Eq.~\eqref{eq:DP}. On the contrary, if $\ell\gg\mathcal{L}$ the randomization process is dominated by the spatial fluctuations in the strength of the spin-orbit coupling \cite{Dugaev_etal}. Within a region of characteristic size $\mathcal{L}$ we have $\delta\sim\mathcal{L}\,\Delta_{BR}/(\hbar v_F)$, and then after a time $t$ we end up with $\delta\left(t\right)\sim\sqrt{t\mathcal{L}/v_F}\times\Delta_{BR}/\hbar$. From the previous arguments we obtain $\tau_s\sim(\Delta_{BR}/\hbar)^2\times\mathcal{L}/v_F$ in that case. 
Typical ripples in graphene extend over $\mathcal{L}=10-50$ nm ($\ll\ell$), so we expect the spin-relaxation times to be determined by geometrical parameters only and not by the scattering times. Diffusion-barrier-induced corrugations in epitaxial growth fronts are also characterized by a constant periodicity, systematically observed in chemical-vapor deposited (CVD) samples. 
Experimentally, however, corrugations do not seem to play an important role in spin transport \cite{avsar_toward_2011}, manifesting basically the same features in CVD graphene as in exfoliated samples.
 
So far, neither of these mechanisms rely on the characteristic Dirac spectrum of low-energy graphene quasiparticles. Numerical studies suggest that the entangled dynamics of the spin and pseudo-spin degree of freedom (the latter associated with the projection of the wave function at different sublattices) leads to fast spin relaxation at energies close to the Dirac point \cite{Van_Tuan_etal}. This mechanism qualitative explains the dependence of $\tau_s$ on the carrier concentrations, which cannot be attributed to the usual EY and DP relations, Eqs.~\eqref{eq:EY}~and~\eqref{eq:DP}.
{\red Experimental validation of the roles of spin and pseudospin degrees of freedom \cite{roche_graphene_2014} and of their relative contributions compared to other sources (e.g. extrinsic, see next section) is crucial to understand the nature of spin relaxation, which is one of the fundamental long-standing puzzles in the way to the full maturity of the field of graphene spintronics.}

\begin{figure*}
\includegraphics[width=1.0\textwidth]{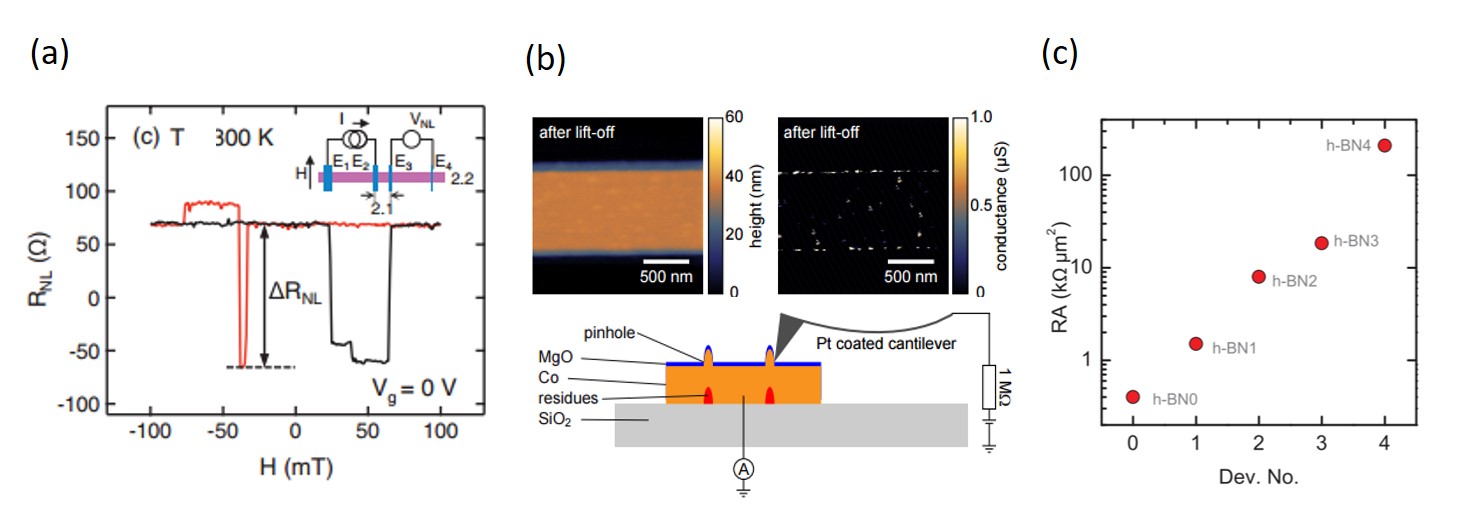}
\caption{
\emph{Advances in Spin Injection contacts.}
(a) Large nonlocal spin valve signal of single layer graphene measured at room temperature. 
Tunnel barrier is a TiO$_2$ seeded MgO layer. From \citet{han_tunneling_2010}. 
(b) Schematic illustration of the conductance force microscopy set-up together with the height and conductance mapping of one of the MgO/Co electrodes.  The bright dots in the conductance scan represent the location of pinholes having conductance of 1~\textmu S. From \citet{drogeler_spin_2016}. 
(c) Tunneling characteristics of hBN barriers measured from several graphene spin valve devices. From \citet{kamalakar_enhanced_2014}.
}
\label{fig:spin-injection}
\end{figure*}

\subsubsection{Extrinsic sources: impurities}


The observed small spin relaxation times in graphene could have an extrinsic origin. \citet{kochan_spin_2014} proposed that local magnetic moments localised on resonant impurities behave as spin hot spots and cause high spin relaxation rates. Based on this mechanism, electrons tend to stay around the impurity \cite{huang_direct_2016} and spins flip is due to the exchange interaction with the magnetic moment on the impurity. Quantitative agreement with experiment on spin relaxation times was achieved even for very low concentration of local moments. It was also predicted that spin relaxation times in single and bilayer graphene have opposite dependences on carrier concentration \cite{kochan_resonant_2015}. This opposite dependence was experimentally verified by carrier concentration dependent spin transport measurements in hBN encapsulated graphene \cite{avsar_electronic_2016}. The latter work also probed a sharp increase of spin relaxation time in bilayer graphene at high carrier densities, which similarly could be explained with the resonant scattering theory. The source of resonant magnetic scatterers is hypothesized to be polymer residues from the device fabrication.

Moreover, Eq.~\eqref{eq:EY} is not consistent with the approximately linear scaling between $\tau_s$ and the diffusion constant at different gate voltages observed in many experiments \cite{jozsa_linear_2009,han_spin_2011,avsar_toward_2011,yang_observation_2011}. The experiments by \citet{jo_spin_2011} are a notable exception for which the relation in Eq.~\eqref{eq:EY} seems to hold. The large spin-orbit coupling required to fit the data ($\Delta\approx10$ meV, close to the intra-atomic coupling in carbon) suggests that  covalently attached adatoms instead of charged impurities dominate spin transport in that case. A unifying theory is provided by \citet{zhang_electron_2012}, suggesting that both the EY and D'yakonov-Perel' mechanisms cohabit in general, their relative strength being dictated by ambient conditions. 
{\purple Recently, \citet{kochan_breakdown_2019} investigated spin relaxation in graphene in proximity to an $s$-wave superconductor in the presence of resonant impurities, which allows comparing the relaxation due to magnetic and spin-orbit impurities. This theoretical work is calling for experimental verification in systems such as superconductivity proximitized graphene \cite{zutic_proximitized_2019}.} 

\subsection{Advances in spin injection contacts}

A tunnel barrier is a key element of graphene spin valve devices to alleviate the conductivity mismatch problem, which dictates that the spin injection efficiency decreases if the contact resistance is too low \cite{rashba_theory_2000}. To achieve a spin dependent tunnel barrier, the pioneering work of \citet{tombros_electronic_2007} utilized an electron beam evaporator to grow $\approx 0.6$~nm thick Al layer on graphene under high vacuum conditions, followed by its oxidation in a 100~mbar O$_2$ pressure.
These initial devices and the follow up works using MgO barriers typically exhibit spin injection efficiencies of $\approx$ 5\% \cite{wang_magnetotransport_2008, yang_observation_2011}. 
On the other hand, \citet{dlubak_are_2010, dlubak_homogeneous_2012}, characterized sputter deposited Al$_2$O$_3$ and MgO layers. Sputter deposition is a standard method to grow barriers in tunnel magnetoresistance (TMR) structures, and resulted in nearly pinhole free $\approx$ 1~nm thick Al$_2$O$_3$ barrier on graphene. Nevertheless, this method destroyed the structural quality of graphene sheets, as evidenced by the observation of strong Raman peaks associated with defects \cite{dlubak_homogeneous_2012}. Spin valves prepared with Al$_2$O$_3$ barriers grown by atomic layer deposition similarly achieved spin injection efficiencies of $\approx 5$\% \cite{yamaguchi_tunnel_2012}. Furthermore, \citet{komatsu_spin_2014} oxidized thermally evaporated yttrium barriers in air, to form 1~nm thick yttrium-oxide tunnel barriers and achieving spin injection efficiency of 15\%. 

Although most spin injection barriers dominated by pinholes, e.g.\ using MgO or Al$_2$O$_3$, allowed spin injection and detection, the extracted short spin relaxation lengths were associated with contact induced spin scattering. Therefore a significant effort has been given to improve the quality of these barriers. 
In this regard, \citet{han_tunneling_2010} achieved tunnel spin contacts with polarization up up to 30\%, by adding a thin Ti seed layer at the interface of MgO and graphene. Spin signals up to 130 \textohm\ were observed, two orders of magnitude higher than the typical values achieved by using transparent contacts, see Fig.~\ref{fig:spin-injection}(a). Consequently, these devices exhibited improved spin relaxation times of up to 0.5~ns. 
Furthermore, \citet{drogeler_spin_2016} increased spin relaxation times up to 12~ns by utilizing MgO barriers evaporated on Cobalt electrodes, just before the mechanical transfer of a encapsulating graphene/hBN heterostructure. Curiously, contact resistance measurements and scanning force microscopy show the presence of pin-holes in the MgO barrier, Fig.~\ref{fig:spin-injection}(b). 

2D crystals such as functionalized graphene or hBN are expected to act as ideal ultrathin tunnel barriers, without pinholes. Prior to spin transport measurements, the potential of hBN as a tunnel barrier was demonstrated by \citet{britnell_electron_2012}. Their results showed that hBN layers fulfill all requirements for a most favorable tunnel barrier \cite{jonsson-akerman_reliability_2000}: 1) It has non-linear but symmetric I-V dependence, 2) such dependence is weakly sensitive to temperature, 3) the resistance of the barrier depends exponentially on thickness, see Fig.~\ref{fig:spin-injection}(c), and 4) it is pin-hole free. 
Reported spin valve devices so far have utilized monolayers of exfoliated or CVD-grown hBN, which have comparable resistance with the graphene channel itself \cite{yamaguchi_electrical_2013, fu_large-scale_2014, kamalakar_enhanced_2014, gurram_spin_2016}. Therefore most of these devices are within the conductivity mismatch range and do not take full advantage of hBN as a tunnel barrier. 
As a different approach, \citet{friedman_homoepitaxial_2014, friedman_hydrogenated_2015} used chemically functionalized graphene as a tunnel barrier. From the nonlocal spin valve measurements, they achieved spin injection efficiencies up to 45\% and 17\% at low temperatures for fluorinated and hydrogenated graphene-based tunnel barriers, respectively. However calculated spin relaxation times in these devices are less than 0.2~ns and the created spin accumulation is very small or even absent at room temperature \cite{friedman_homoepitaxial_2016}. 
Recently, \citet{gurram_bias_2017} have observed bias-induced enhanced spin injection and detection efficiencies up to 100\% in tunnel junctions with bilayer hBN barriers. 
Similar results were obtained for trilayer hBN tunnel barriers \cite{leutenantsmeyer_efficient_2018}. These results are not yet understood.

\begin{figure*}
\includegraphics[width=1.0\textwidth]{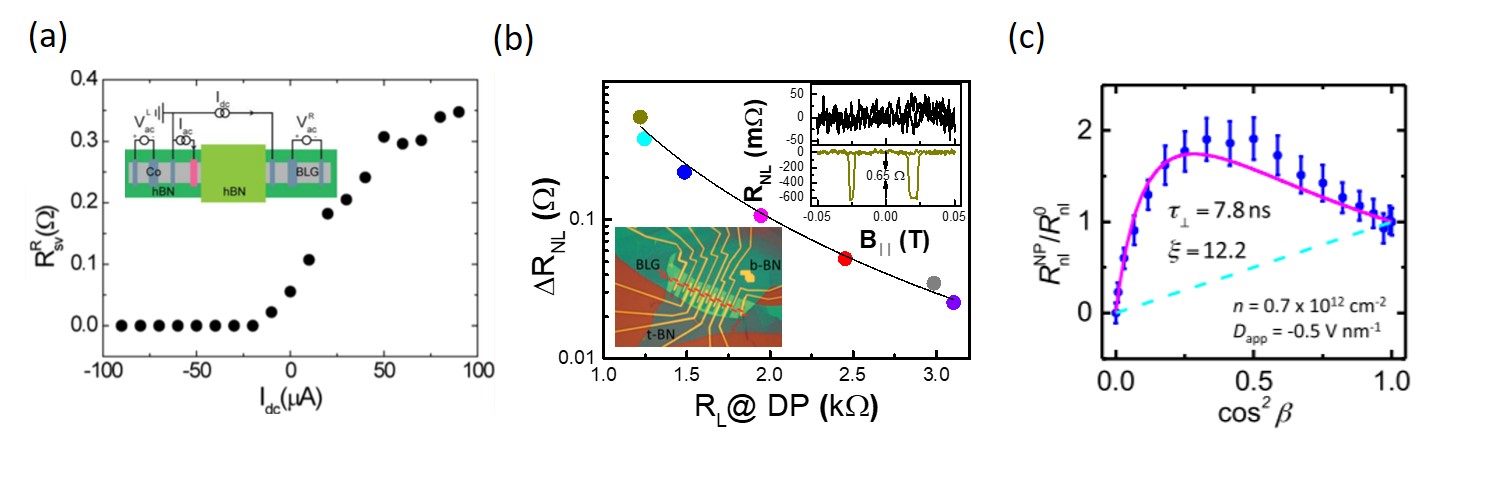}
\caption{
\emph{Electric-Field Dependent Spin Transport in 2D-based Heterostructures.}
(a) The amplitude of the detected spin signal as a function of drift current, demonstrating spin drift. Inset shows the measurement geometry. From \citet{ingla-aynes_eighty-eight_2016}. 
(b) Dependence of nonlocal spin signal in bilayer graphene on the local resistance at the Dirac point. The latter increases with the vertical electric field (from yellow to purple dot). At higher fields, the spin current is switched off by the gapped channel. Top inset shows spin valve measurements at zero (dark yellow line) and high (black line) displacement fields, demonstrating a spin switch effect. From \citet{avsar_electronic_2016}. 
{\red (c) Oblique Hanle measurements near the charge neutrality point in a bilayer graphene spin valve, for different angle $\beta$ relative to the graphene plane. The magenta curve is a fit to the experimental data (blue points) and reveals an anisotropy of $\sim 12$ in the spin relaxation time. Away from the charge neutrality point, the anisotropy diminishes and the response becomes similar to the cyan dashed line. From \citet{xu_strong_2018}.}
}
\label{fig:spinFET}
\end{figure*}

\subsection{Alternative spin injection and detection techniques}
{\red Besides the use of electrical and optical means, ferromagnetic resonance spin pumping has also been utilized to inject spins into graphene. In this technique, a ferromagnet is placed in contact with graphene and the magnetization of the ferromagnet is driven into resonant precession by the use of an oscillating magnetic field. The angular momentum transfer from the precessing ferromagnet drives a spin current into graphene, while this spin transfer is observed as an enhanced damping of the ferromagnetic resonance. Different than the electrical spin injection method, this dynamical method does not require any tunnel barrier as it allows injection via transparent interfaces. \cite{patra_dynamic_2012} reported the first evidence of spin pumping into chemically grown graphene by demonstrating a remarkable broadening of the ferromagnetic resonance absorption peak in Py/graphene and Co/graphene films. The significant absorption of angular momentum suggested that SOC strength in proximitized CVD graphene is larger than for pristine graphene \cite{patra_dynamic_2012, singh_spin_2013}. Still, the exclusive use of ferromagnetic broadening to quantify spin injection was cautioned by the work of \citet{berger_magnetization_2014}, which demonstrated that cobalt grown on graphene has enhanced magnetic disorder and hence the observed broadening may have a contribution from the structural change of the ferromagnet. This uncertainty was addressed by the work of \citet{tang_dynamically_2013}, which unambiguously proved dynamical spin injection at room temperature by detecting the injected spin current via the inverse SHE (ISHE) in Pd electrodes. A similar approach was later shown using the ISHE of graphene itself \cite{ohshima_observation_2014}. Similar to ISHE, there are other methods for electrically probing spin transport in graphene. Detailed quantum interference measurements, combining weak localization and universal conductance fluctuations, probed decoherence in graphene \cite{lundeberg_defect-mediated_2013} and indicated that magnetic defects are the dominating source of spin scattering. This identification of an extrinsic source of spin relaxation in the form of magnetic defects is in agreement with later theoretical predictions \cite{kochan_resonant_2015} and spin valve measurements \cite{avsar_electronic_2016}.

The unique electronic structure of graphene paves the way for unconventional ways to detect and create spin transport, even without using magnetic contacts. Under a large applied magnetic field, graphene devices in a Hall bar geometry exhibit a giant nonlocal resistance \cite{abanin_giant_2011}. This phenomenology, similar to that of the SHE, was attributed to the creation of spin-up electrons and spin-down holes due to Zeeman splitting near the charge neutrality point, without any role of SOC \cite{ abanin_giant_2011-1}. A further work indicated that thermoelectric effects, i.e. the interaction between thermal and electronic transport, could have a contribution to this nonlocal signal as well \cite{renard_origins_2014}. The insight that thermoelectric transport is interrelated with spin transport has developed in recent years into the active subfield of spin caloritronics \cite{bauer_spin_2012}. This interrelation was demonstrated in graphene via the detection of spin currents without the use of magnetic contacts \cite{vera-marun_nonlinear_2012}, where the spin-to-charge conversion mechanism relies on the energy-dependent transmission of graphene, the same principle as that of the thermoelectric Seebeck effect \cite{vera-marun_nonlinear_2011}. Such a spin-to-charge conversion proved to be a general phenomenon, advantageous for detecting spin transport in mesoscopic systems \cite{stano_nonlinear_2012, yang_spin-current_2014, Rameshti2015}. More recently, the role of thermal gradients on the modulation of spin signals has also been demonstrated \cite{sierra_thermoelectric_2018}. Graphene and related 2D materials are fundamentally interesting systems for the exploration of novel spin injection and detection approaches, in particular within the still young subfield of 2D spin caloritronics.} 
{\purple Finally, one-dimensional ferromagnetic edge contacts to graphene have started to be explored for spin injection and detection \cite{karpiak_1d_2017, xu_spin_2018, guarochico_spintronics_2017}. 
In this regard, graphene in direct contact to a ferromagnet enables the study of an electrically-tunable magnetic proximity induced spin polarisation \cite{lazic_effective_2016, asshoff_magnetoresistance_2017}, which is a promising functionality for future graphene-based spin logic devices \cite{dery_nanospintronics_2012}.}  

\subsection{Vertical junctions for spin memories}
Several 2D crystals, including graphene, hBN and Mo$S_2$, have been integrated into vertical GMR and TMR structures as non-magnetic barriers \cite{cobas_graphene_2012, cobas_graphene-based_2013, chen_layer-by-layer_2013, iqbal_spin_2013, meng_vertical_2013, godel_voltage-controlled_2014, li_magnetic_2014, wang_spin-valve_2015, dankert_tunnel_2014}. 
These structures are the most prominent class of spintronic devices, widely used as magnetic
sensors. 
Other uses include graphene as a barrier for spin injection into semiconductors \cite{erve_low-resistance_2012}, to prevent oxidation of ferromagnetic electrodes \cite{dlubak_graphene-passivated_2012}, or as a spin-conserving channel in a hot-electron transistor 
\cite{banerjee_spin_2010}.

Interest towards the integration of 2D crystals in vertical structures was driven by the prediction that few-layer graphene could act as a perfect spin filter \cite{karpan_graphite_2007, karpan_theoretical_2008, yazyev_magnetoresistive_2009}. 
Improved advancements on device fabrication have allowed to study the nature of magnetoresistance in graphene and hBN based vertical devices, evidencing the role of charge transfer and proximity effects. The latter include the spin-splitting induced in graphene due to proximity to the ferromagnetic electrodes \cite{asshoff_magnetoresistance_2017}, and to identify the role of individual resonant defect states in hBN to enhance TMR \cite{asshoff_magnetoresistance_2018}. All these approaches set the prospect of engineering vertical magnetoresistance in 2D-based junctions, which are the key elements within current memory architectures of magnetic random access memories \cite{parkin_magnetically_2003}.

\begin{figure*}
\includegraphics[width=1.0\textwidth]{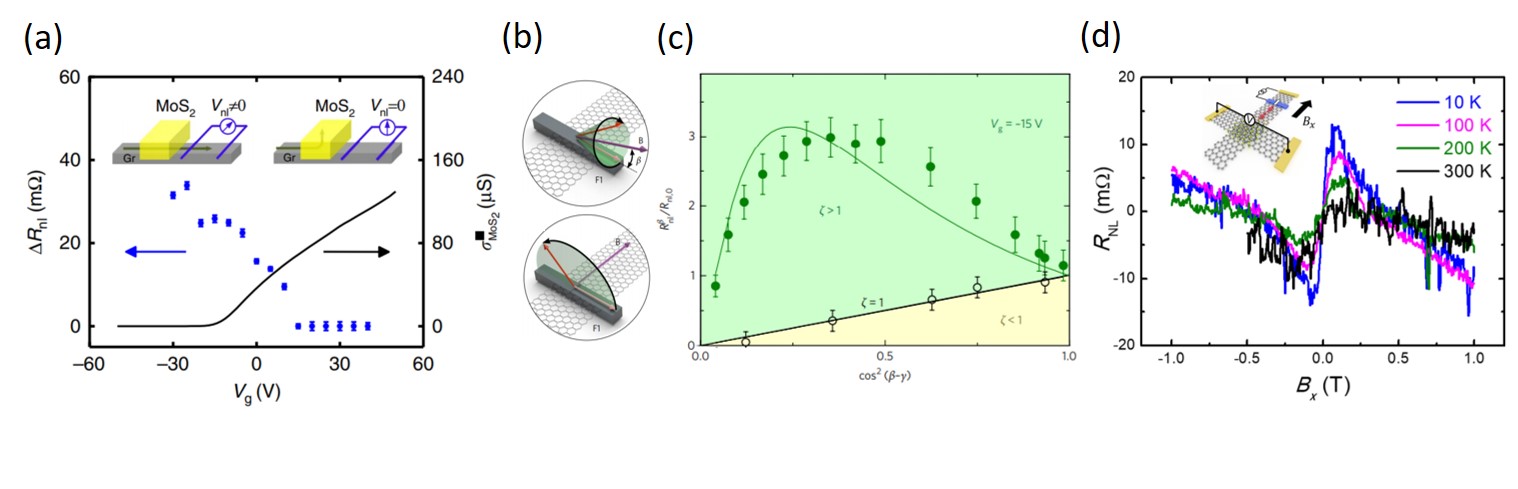}
\caption{
\emph{Proximity-enabled graphene/TMDC devices.}
{\red (a) Nonlocal spin signal and conductivity for graphene/MoS$_2$ device, demonstrating a field-controlled spin absorption. From \citet{yan_two-dimensional_2016}. (b) Top schematic shows the spin precession cone under an oblique magnetic field, applied in a plane that contains the easy axis of the ferromagnetic electrodes and that is perpendicular to the substrate. Bottom schematic shows the spin precession with magnetic field applied in plane and perpendicular to the easy axis of ferromagnets. Here, the spin precesses on a plane perpendicular to the substrate. From \citet{benitez_strongly_2018}. (c) Anisotropy measurements in a graphene/WS$_2$ spin valve device. Solid green dots represent the measurements taken at V$_BG$ = -15V, with the green line a fit revealing an anisotropy of $\sim 10$. Open black dot shows data acquired in a reference graphene spin valve without WS$_2$, lacking any anisotropy. From \citet{benitez_strongly_2018}. (d) Spin-to-charge conversion in a Hall bar detector within a graphene/MoS$_2$ heterostructure, at different temperatures. The inset shows the measurement configuration. From \citet{safeer_room-temperature_2019}.}
}
\label{fig:proximity}
\end{figure*}

\subsection{Electric-field effect in mono- and bilayer-graphene}

As discussed previously, hBN has also been employed  to develop device concepts which could pave the realization of new types of spin-based logic transistors. 
In a dual-gated device architecture utilizing hBN as a dielectric layer, \citet{ingla-aynes_eighty-eight_2016} controlled the propagation of spin current at room temperature using the drift of electron spins, as shown in Fig.~\ref{fig:spinFET}(a). Here, the spin relaxation length of 7.7~\textmu m, was demonstrated to be controlled from 0.6~\textmu m up to 90~\textmu m when a DC current of $\pm 90$~\textmu A was applied. Depending on the polarity of the DC current, directional control of spin transport was achieved. This spin drift effect demonstrated how longitudinal electric fields can guide and extend the propagation of spin current. 

A similar dual-gated structure, this time using bilayer graphene, was also employed to study a different phenomenon: the role of perpendicular electric fields. Here a transport gap was induced on bilayer graphene by the application of a vertical electric field at low temperature, which allowed a purely electrostatic control of spin current propagation \cite{avsar_electronic_2016}. 
In an entirely opposite trend to the device resistance, the spin signal near the charge neutrality point was observed to rapidly decrease as the displacement field was increased, and eventually the signal becomes undetectable. 
This device therefore served as an spin-transport analogy to the charge-based field-effect transistor, demonstrating a full electrostatic-gate control of spin current propagation, see Fig.~\ref{fig:spinFET}(b). 
{\red Furthermore, the unique spin-valley coupled band structure of bilayer graphene in such dual-gated device architecture enables the exciting possibility of modulating the spin-lifetime anisotropy. By performing Hanle spin precession measurements under obliquely oriented magnetic fields, \citet{xu_strong_2018} demonstrated that the ratio between out-of-plane (7.8 ns) and in-plane (0.64 ns) spin lifetimes is as large as $\sim 12$ near the charge neutrality point, as shown in Fig. 8(c). Such a large anisotropy is a result of an out-of-plane spin orbit field originated from the induced- band gap in bilayer graphene. When increasing the carrier concentration under a fixed vertical displacement field, both in-plane and out-of-plane spin lifetimes eventually become comparable and therefore the anisotropy disappears. These groundbreaking results, which offer a novel approach to manipulating spin information, were independently demonstrated by \citet{leutenantsmeyer_observation_2018} as well.}

\begin{figure*}
\includegraphics[width=1.0\textwidth]{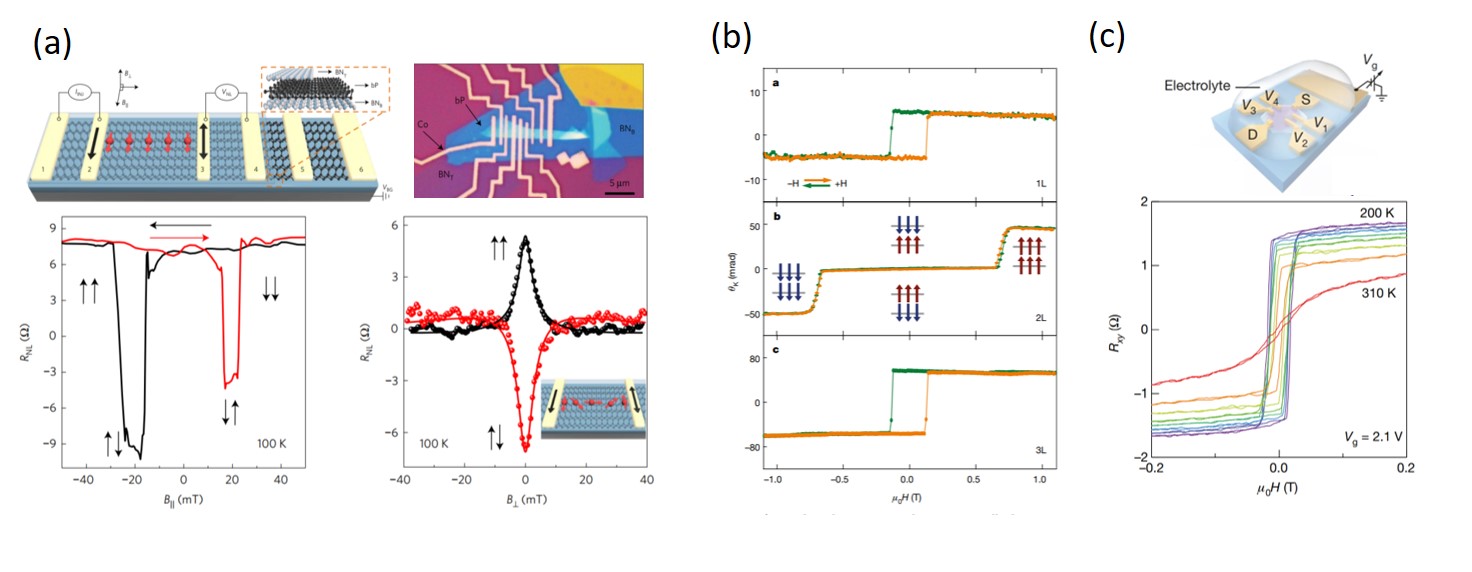}
\caption{
\emph{Spintronics in 2D Materials Beyond Graphene.}
(a) Spin transport in black phosphorus spin valves. Top panels show the device schematics and the optical image of a completed device, respectively. Bottom panels show spin valve and spin precession curves taken at 100~K, respectively. From \citet{avsar_gate-tunable_2017}. 
(b) Polar magneto-optical measurement for CrI$_3$, showing hysteresis for ferromagnetic monolayer (1L) and trilayer (3L), and vanishing Kerr rotation for antiferromagnetic bilayer (2L). From \citet{huang_layer-dependent_2017}. 
(c) Upper panel represents the device geometry, where electrolyte gating technique was employed to induce an electric field. This tunes the density of states in magnetic Fe$_3$GeTe$_2$. Bottom panel shows the anomalous hall effect measurements as a function of temperature in a four-layers-thick Fe$_3$GeTe$_2$. From \cite{deng_gate-tunable_2018}
}
\label{fig:alternative2}
\end{figure*}

\subsection{Proximity-enabled graphene/TMDC novel devices}
{\red The ability to control both the flow of spin information and the anisotropy of spin lifetime via an electrostatic gate in bilayer graphene relies on the presence of its band gap, which is only evident in experiments at low temperature. On the other hand, similar electrostatic control of spin information has been recently realized at room temperature, by exploring graphene/TMDC heterostructures. Electric field-controlled spin current in graphene/MoS$_2$ spin valve devices was first reported as a result of gate-tunable spin absorption into the adjacent MoS$_2$ layer \cite{dankert_electrical_2017, yan_two-dimensional_2016}, see Fig.~9 (a). Later, two independent groups demonstrated that strong spin-valley coupling in WS$_2$ or MoS$_2$ results in a change of over one order of magnitude between the spin lifetimes for in-plane and out-of-plane spins \cite{ghiasi_large_2017, benitez_strongly_2018}, see Fig.~9 (b-c), in good agreement with earlier theoretical predictions \cite{cummings_giant_2017}. These encouraging results in heterostructure systems enabled further theoretical investigations focusing on converting charge current into spin current, either using the Spin Hall effect (SHE) which can be induced via proximity in graphene \cite{garcia_spin_2017, milletari_covariant_2017} or the Rashba-Edelstein effect (REE) present at the graphene/TMDC interface \cite{offidani_optimal_2017}. 
The corresponding inverse effects (ISHE and IREE) are also possible, leading to reciprocal spin to charge conversion. 

Towards this direction, \citet{safeer_room-temperature_2019} studied graphene/MoS$_2$ heterostructures and demonstrated the presence of a proximity-induced ISHE in graphene, with an additional spin to charge conversion mechanism that could be indistinguishably attributed to either a proximity-induced IREE in graphene or ISHE within the MoS$_2$ layer (Fig.~9(d)). Soon after, charge-to-spin conversion due to REE in a graphene/monolayer WS$_2$ heterostructure was demonstrated, also evidencing its carrier density and temperature dependences \cite{ghiasi_charge--spin_2019, benitez_tunable_2019}.} 
{\purple Similarly, REE at room temperature was also observed in graphene/TaS$_2$ \cite{li_electrical_2019} and graphene/MoTe$_2$ \cite{hoque_all-electrical_2019} heterostructures, with these works also demonstrating the carrier density dependence of the REE efficiency and polarity in agreement with theoretical predictions \cite{offidani_optimal_2017}.} 
These recent experiments unambiguously prove charge-to-spin conversion via proximity-induced effects (SHE, REE), by combining Hall probes with standard ferromagnetic electrodes within the same device architecture. The latter is a critical advancement compared to the earlier proximity studies, which utilized only Hall bar (non-magnetic) electrodes \cite{avsar_spinorbit_2014, wang_proximity-induced_2015}.

\subsection{Spintronics with 2D semiconductors and 2D magnets}

Spintronics in 2D semiconductor materials such as black phosphorus (BP) and TMDCs can offer functinality that is not possible by using only graphene, for example gate controlled amplification/switching actions. Towards this, \citet{avsar_gate-tunable_2017} demonstrated all electrical spin injection, transport, manipulation and detection in ultra-thin BP-based spin valves at room temperature. Based on four-terminal Hanle spin precession measurements, spin relaxation times up to 4.5~ns with spin relaxation lengths exceeding 6~\textmu m were extracted, see Fig.~\ref{fig:alternative2}(a). Temperature and gate voltage dependences for spin and momentum relaxation times were in good agreement with first-principles calculations, showing that Elliott-Yafet relaxation is dominant in BP \cite{li_electrons_2014, kurpas_spin-orbit_2016}. Further work is required to explore other spintronic properties, like directional spin transport in this anisotropic crystal. 

Initial efforts to investigate magnetism in 2D materials mostly focused on proximity induced \cite{wang_proximity-induced_2015} and defect induced \cite{yazyev_defect-induced_2007} magnetism. Based on the Mermin-Wagner theorem \cite{mermin_absence_1966}, intrinsic ferromagnetic order is indeed not expected in 2D single layers. This was initially supported by the experimental results obtained from 2D Cr$_2$Ge$_2$Te$_6$ where magnetism is absent in its monolayer, at least down to the lowest studied temperatures (4.7~K) \cite{gong_discovery_2017}. In this material ferromagnetic order persists nonetheless down to bilayer, but the critical temperature significantly decreases from 68~K (bulk) to 30~K (bilayer), which can be explained by the thermal excitation of spin waves.

On the other hand, the Mermin-Wagner restriction can be lifted if a material displays strong SOC and magnetocrystalline anisotropy. As shown in Fig.~\ref{fig:alternative2}(b), layer dependent magnetic ordering in CrI$_3$ was demonstrated down to monolayer thickness, with its spins constrained to lie vertical to the lattice plane as described by the Ising model \cite{huang_layer-dependent_2017}. 
The layered structure of CrI$_3$ provides interesting opportunities: an odd number of layers results in long range ferromagnetic ordering, while an even number are antiferromagnetic. In the bilayer case, application of an external magnetic field above the coercive field makes the material ferromagnetic; all layers are aligned to the same directions. By taking advantage of this polarization switch, giant tunneling magnetoresistance was demonstrated in vertically studied samples \cite{klein_probing_2018, song_giant_2018, wang_very_2018}. At constant voltage bias, such a magnetoresistance changes the current by nearly one million percent if thicker CrI$_3$ is employed \cite{kim_one_2018}. It was also demonstrated that the critical field required to switch from antiferromagnetism to ferromagnetism depends on the vertically applied electric field, see Fig.~\ref{fig:alternative2}(c) \cite{jiang_controlling_2018, huang_electrical_2018}. 

Demonstration of 2D magnetism in Cr$_2$Ge$_2$Te$_6$ and CrI$_3$ has pushed a rapid progress towards the discovery of new 2D magnetic materials, such as CrBr$_3$ \cite{ghazaryan_magnon-assisted_2018}, Fe$_3$GeTe$_2$ \cite{deng_gate-tunable_2018} and  VSe$_2$ \cite{bonilla_strong_2018}. Based on magneto-optical measurements, it was shown that VSe$_2$ shows ferromagnetic ordering persisting up to room temperature in its monolayer, in sharp contrast to its bulk counterpart, which is paramagnetic. 
Similarly, room temperature magnetic ordering was discovered in metallic Fe$_3$GeTe$_2$ crystals by employing the ionic gating technique. 
{\red The use of room-temperature 2D magnets for potential spintronic applications is further enabled by their growth using molecular beam epitaxy techniques down to monolayer thickness, as demonstrated for MnSe$_2$ \cite{ohara_room_2018}. Room temperature spintronic devices could integrate the spin transport properties of graphene with the memory of the 2D magnet, exploiting proximity-induced magnetization by the 2D magnet into graphene as demonstrated via the proximity anisotropic magnetoresistance effect in graphene/CrB$_3$ heterostructures \cite{ghazaryan_magnon-assisted_2018}. Recent observation of defect-induced long-range magnetism in air-stable, metallic PtSe$_2$ \cite{avsar_defect_2019} and semiconducting MoTe$_2$ \cite{guguchia_magnetism_2018} expands the range of 2D magnets into materials that are intrinsically nonmagnetic and would otherwise be overlooked.}
The emerging mix of experimental techniques and new materials, and the resulting new effects like gate-tunable magnetism, creates exciting prospects for 2D-based voltage-controlled magnetoelectronics.

\section{Final remarks}
{\red Since the first demonstration of spin transport in graphene \cite{tombros_electronic_2007} we have seen a remarkable progress in 2D-based spintronics, including graphene related materials, getting tantalisingly close towards practical applications. 
{\purple This includes, but is not limited to: a two orders of magnitude enhancement in graphene spin lifetime \cite{drogeler_spin_2016}, high quality spin transport in large-scale graphene fabricated on conventional Si/SiO$_2$ \cite{avsar_toward_2011, kamalakar_long_2015, gebeyehu_spin_2019} and flexible polymer \cite{serrano_two-dimensional_2019} substrates, gate-induced spin manipulation in 2D semiconductors \cite{avsar_gate-tunable_2017} and graphene/TMDC heterostructures \cite{yan_two-dimensional_2016, dankert_electrical_2017}, proximity induced spin-to-charge conversion in graphene/TMDC heterostructures \cite{safeer_room-temperature_2019, li_electrical_2019, ghiasi_charge--spin_2019}, and the recent discovery of 2D magnets that can even remain magnetically ordered down to a few atoms thick at room temperature \cite{deng_gate-tunable_2018}.} 

However, there are long-standing challenges that need to be addressed for exploiting the full potential of 2D spintronics applications. There is still room for improving spin lifetime and spin relaxation length by studying fully encapsulated, high mobility graphene spin valves. We believe there are two major breakthroughs expected when these devices reach the electronic quality of state of art charge-based devices. 
First, experimental results in such high-quality spintronic devices will guide the theory to convincingly identify the limiting spin relaxation mechanisms. 
Second, analogous to  the demonstration of micrometer-scale ballistic charge transport at room temperature \cite{mayorov_micrometer-scale_2011}, the realization of ballistic spin transport at room temperature could open a whole new range of opportunities. 
{\purple The latter relates to both applications (e.g.\ enabled by spin pumping \cite{Bercioux2012} or spin filtering \cite{Valli2018} effects), and to developing a theoretical description to interpret the role of device geometry, spin precession, and transport in a regime beyond the diffusive approximation \cite{Bercioux2010, vila_nonlocal_2019}.} 
Considering the field started only over a decade ago, and its current progress, the vision of low-power all-spin-logic devices based on 2D van der Waal heterostructures seems within the reach of the next decade.}

\begin{acknowledgments}

{\purple We thank A.\ Ferreira, D.\ Bercioux, E.\ Sherman, S.\ O.\ Valenzuela, A.\ G.\ Moghaddam, M.\ V.\ Kamalakar, I.\ {\v Z}uti{\'c}, J.\ Fabian and S.\ Roche for valuable discussions.} 
We acknowledge the financial support from the 
European Union's Horizon 2020 research and innovation program under Grant Agreement Nos.\ 696656 and 785219 (Graphene Flagship Core 1 and 2). 
BJvW acknowledges the support by the Zernike Institute for Advanced Materials, and the NWO Spinoza prize awarded by the Netherlands Organisation for Scientific Research (NWO). 
{\red B.\"O.\ acknowledges support by the National Research Foundation, Prime Minister's Office, Singapore under its Investigator Award (Project No. NRF-NRFI2018-08) and its Medium Sized Centre Program.}
IJVM acknowledges the support of the Future and Emerging Technologies (FET) Programme within the Seventh Framework Programme for Research of the European Commission, under FET-Open Grant No.\ 618083 (CNTQC).

\end{acknowledgments}

\bibliography{theory_edited,Zotero_v02,Review-extras}

\end{document}